\documentclass[runningheads]{arxiv_llncs}
 
\usepackage[final]{arxiv}
\usepackage{appendix}

\usepackage{arxivabbrv}

\usepackage{graphicx}
\usepackage{booktabs}
\usepackage{multirow}
\usepackage[accsupp]{axessibility}  

\usepackage[pagebackref,breaklinks,colorlinks]{hyperref}

\usepackage{orcidlink}

\begin{document}

\title{ConvBench: A Multi-Turn Conversation Evaluation Benchmark with Hierarchical Capability for Large Vision-Language Models} 


\newcommand{\blue}[1]{\textcolor[rgb]{0.00,0.00,1.00}{#1}}
\newcommand{\red}[1]{\textcolor[rgb]{1.00,0.00,0.00}{#1}}

\author{Shuo Liu\inst{1} \and
Kaining Ying\inst{1,2} \and
Hao Zhang\inst{1,3} \and
Yue Yang\inst{1,4}  \and
Yuqi Lin\inst{1,2} \and
Tianle Zhang\inst{1,5} \and
Chuanhao Li\inst{1,6} \and
Yu Qiao\inst{1} \and
Ping Luo\inst{1,7} \and
Wenqi Shao\inst{1}$^{\star}$
  \and
Kaipeng Zhang\inst{1}\thanks{Equal Corresponding Authors}}

\authorrunning{S. Liu et al.}

\institute{Shanghai Artificial Intelligence Laboratory, Shanghai, China \and
Zhejiang University, Hangzhou, China \and
Xi'an Jiao Tong University, Xi'an, China \and
Shanghai Jiao Tong University, Shanghai, China \and
The University of Electronic Science and Technology of China, Chengdu, China \and
Beijing Institute of Technology, Beijing, China \and
The Chinese University of HongKong, HongKong, China \\
\email{liushuo@pjlab.org.cn, shaowenqi@pjlab.org.cn, zhangkaipeng@pjlab.org.cn}\\} 

\maketitle

\begin{abstract}
This paper presents ConvBench, a novel multi-turn conversation evaluation benchmark tailored for Large Vision-Language Models (LVLMs). Unlike existing benchmarks that assess individual capabilities in single-turn dialogues, ConvBench adopts a three-level multimodal capability hierarchy, mimicking human cognitive processes by stacking up perception, reasoning, and creativity. Each level focuses on a distinct capability, mirroring the cognitive progression from basic perception to logical reasoning and ultimately to advanced creativity. ConvBench comprises 577 meticulously curated multi-turn conversations encompassing 215 tasks reflective of real-world demands. Automatic evaluations quantify response performance at each turn and overall conversation level. Leveraging the capability hierarchy, ConvBench enables precise attribution of conversation mistakes to specific levels. Experimental results reveal a performance gap between multi-modal models, including GPT4-V, and human performance in multi-turn conversations. Additionally, weak fine-grained perception in multi-modal models contributes to reasoning and creation failures. ConvBench serves as a catalyst for further research aimed at enhancing visual dialogues.

\end{abstract}

\begin{figure}[t]
  \centering
  \includegraphics[width=\textwidth]{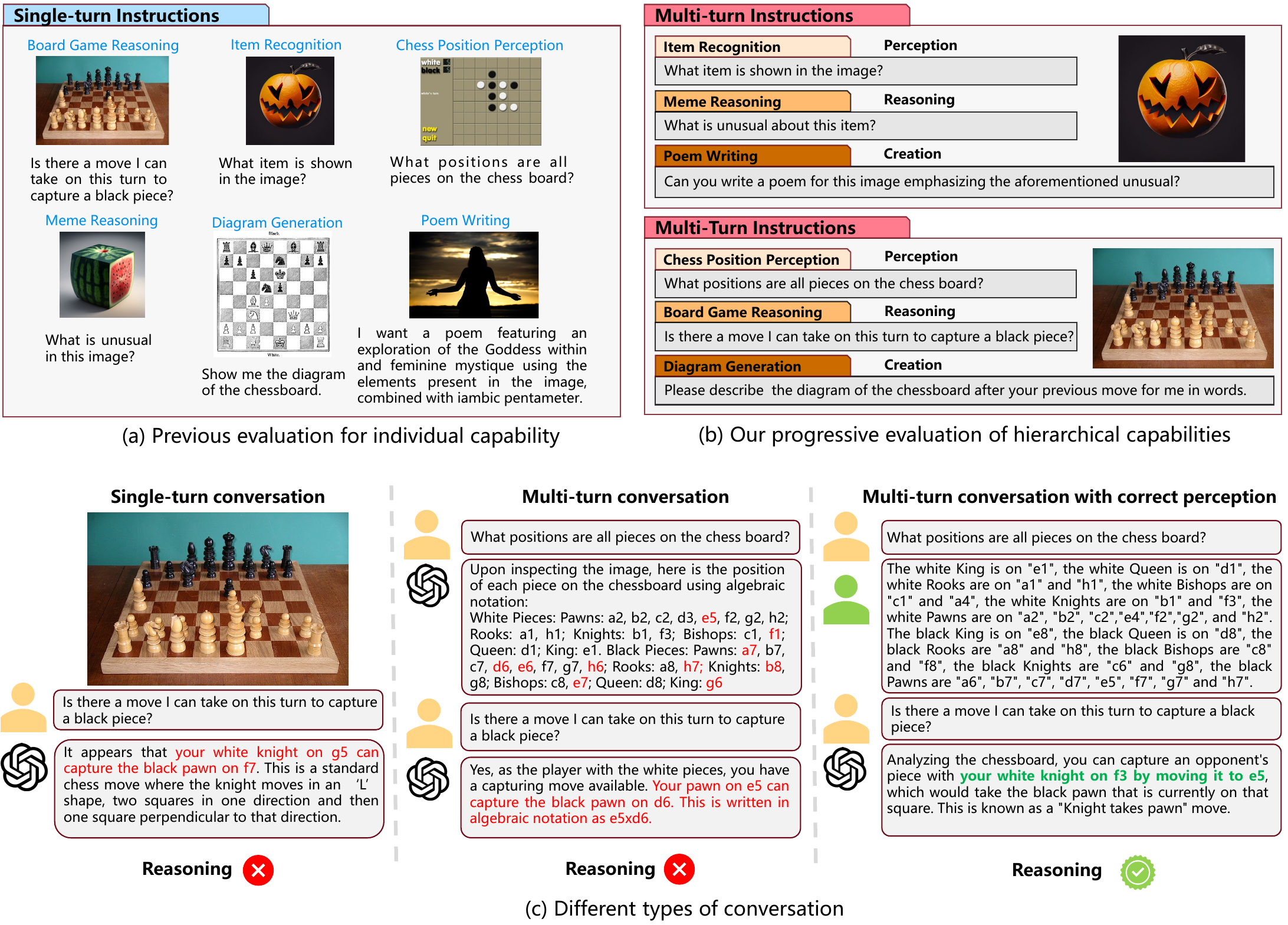} 
  \caption{The comparison between previous evaluation benchmarks (a) and our ConvBench (b). Previous benchmarks assess capabilities independently in a single-turn conversation, while our ConvBench evaluates multi-turn conversation by progressively assessing perception, reasoning, and creativity. (c) shows that multi-turn conversations with hierarchical capabilities can easily attribute mistakes to perception errors.}
  \label{fig:teaser}
\end{figure}

\section{Introduction}
\label{sec:intro}

Large Vision-Language Models (LVLMs)~\cite{GPT4V, Gemini, Liu2023VisualIT, Dong2024InternLMXComposer2MF}, have demonstrated remarkable success in various multimodal applications such as open-world visual question answering~\cite{Liu2023VisualIT, Gao2023LLaMAAdapterVP}, visual dialogue~\cite{Ye2023mPLUGOwlME}, and medical service~\cite{moor2023medflamingo, li2023llavamed}. Due to the great potential in advancing artificial general intelligence, a tremendous surge of research activity has been devoted to improving the performance of LVLMs by various techniques, including effective training strategies~\cite{hu2021lora, Li2023MonkeyIR, Dong2024InternLMXComposer2MF, Liu2023VisualIT}, high-quality image-text dataset~\cite{Dong2024InternLMXComposer2MF, liu2023improved, Chen2023ShareGPT4VIL}, robust model architectures~\cite{liu2023improved, Gao2023LLaMAAdapterVP}. However, previous benchmarks such as VQAv2~\cite{balanced_vqa_v2} and COCO Captioning~\cite{chen2015microsoft} are not enough to assess the performance of these LVLM models, which are designed to solve user's general-purpose requests. Therefore, it is urgent to build a challenging evaluation benchmark to measure the advancement of LVLMs.


Recent studies~\cite{Yu2023MMVetEL, Liu2023MMBenchIY, fu2023mme, xu2023lvlm} have tackled this challenge by introducing a variety of evaluation benchmarks modeled on visual question-answering. Notably, LVLM-eHub~\cite{xu2023lvlm}, MME \cite{fu2023mme}, SEED-Bench \cite{Li2023SEEDBenchBM}, and MMBench \cite{Liu2023MMBenchIY} have amassed numerous test samples to evaluate key multimodal capabilities such as perception and reasoning. Despite their simplicity, these benchmarks can reveal the downside of current LVLMs. In addition, VisIT-Bench \cite{Bitton2023VisITBenchAB} measures how LVLMs respond to various real-world requests by collecting a broad spectrum of tasks from humans. Mathvista \cite{Lu2023MathVistaEM} and MMMU \cite{Yue2023MMMUAM} assess the reasoning and comprehension on expert-level domain knowledge.

However, the above benchmarks \cite{xu2023lvlm, fu2023mme, Lu2023MathVistaEM, Yue2023MMMUAM} evaluate each capability dimension independently in a single-turn conversation as shown in Figure~\ref{fig:teaser} (a), which suffers from two limitations. \textit{i) Naive capability hierarchy.} Previous benchmarks treat different multimodal capabilities independently while ignoring the fact that multimodal capabilities are highly dependent on each other. Evaluating different capabilities independently would make it hard to conduct error attribution. For example, When the model gives the wrong response to a reasoning question, it is unclear whether it is attributed to the perception or reasoning error of LVLMs (see Figure~\ref{fig:teaser} (c)). \textit{ii) Disparity in human preference.} Multi-turn conversation and instruction-following ability are critical elements for human preference. Most previous benchmarks for LVLMs focus on single-turn tests with one instruction and response. However, multi-turn dialogue is the more likely way for general-purpose assistants to collaborate with humans to solve diverse tasks. 


To tackle these challenges, we introduce ConvBench, a benchmark designed for multi-turn conversation evaluation that progressively examines the perception, reasoning, and creativity capabilities of LVLMs.  Inspired by the fact that humans reason based on what they perceive and generate new ideas through a combination of perceptual and reasoning skills, we build a three-level hierarchy of multimodal capabilities ranging from perception to reasoning and finally to creativity. Such a capability hierarchy is instantiated for each test sample within a framework of multi-turn dialogue, as shown in Figure~\ref{fig:teaser}(b). Throughout the conversation, LVLMs are tasked with progressively addressing challenges across these three levels. Evaluating LVLMs using a multi-turn conversation framework with capability hierarchy enables error attribution as shown in Figure~\ref{fig:teaser}(c). 

To establish the capability hierarchy, each instance in ConvBench is composed of an input image, three progressive instructions for the three-level capability hierarchy, three human-verified references, and an instruction-conditioned caption verified by humans. Specifically, the annotation starts by creating three progressive instructions based on an input image in a multi-turn manner. Initially,  we curate the perception instruction for the first level, followed by the reasoning and creativity instructions, which are generated  in response to the instructions and answers at preceding levels. We then annotate image captions tailored for these instructions following~\cite{Bitton2023VisITBenchAB}. GPT-4V, with the help of instruction-conditioned captions, produces preliminary outputs by feeding it input images and instructions. A subsequent human verification step is employed to  ensure the high quality of the reference responses.

Overall, ConvBench comprises $577$ meticulously curated multi-turn QA samples, spanning $71$, $65$, and $79$ distinct types of perception, reasoning, and creation tasks, as depicted in Figure~\ref{fig:tasks}, respectively. We assess $19$ publicly available LVLMs, including the advanced GPT4-V \cite{GPT4V}, employing various assessment methods such as direct grading and pairwise comparison.  The evaluation is conducted in an ablative manner, enabling error attribution, as illustrated in Figure~\ref{fig:teaser}(c). The evaluation results reveal several innovative findings: i) Our ConvBench poses a substantial challenge for current LVLMs, notably GPT4-V \cite{GPT4V}, which only achieves $39.51$\% overall score in pairwise evaluation. ii) Through extensive ablative evaluation, we conclude that weak perception capability undermines LVLMs’ reasoning and creativity and limited reasoning capacity also hinders creativity. iii) LVLMs demonstrate weak performance in perception, particularly in fine-grained recognition, object detection, and tasks related to detailed descriptions.

The contributions of this work are summarized as follows: Firstly, we introduce ConvBench, a multi-turn conversation evaluation benchmark, to assess various LVLMs. ConvBench comprises 577 meticulously curated test samples, covering a wide range of multimodal tasks. Secondly, ConvBench is built upon a three-level hierarchy of multimodal capabilities from basic perception to intricate reasoning, and to advanced creativity. This capability hierarchy facilitates error attribution. Thirdly, our extensive evaluation of diverse LVLMs reveals that ConvBench represents a significant challenge for current LVLMs. For instance, even the advanced \cite{GPT4V} only achieves $39.51$\%, $38.47$\%, $39.34$\%, and $37.61$\% overall conversation, perception, reasoning, and creativity score, respectively,  in pairwise evaluation method.   

\section{Related Work}
\subsubsection{Large Vision-Language Models.} Building upon the achievements of Large Language Models (LLMs)~\cite{Chatgpt, Chung2022ScalingIL, Touvron2023LLaMAOA}, Large Vision-Language Models (LVLMs) ~\cite{GPT4V, Zhu2023MiniGPT4EV, Liu2023VisualIT, Qwen-VL, Ye2023mPLUGOwlME, Chen2023InternVLSU, Li2023OtterAM, Gemini, Dong2024InternLMXComposer2MF, Gao2023LLaMAAdapterVP} have recently showcased remarkable proficiency across a variety of tasks, demonstrating advanced perception, reasoning, and creative capabilities. A favored approach to enhancing LVLMs involves integrating visual knowledge into the semantic framework of LLMs, thereby leveraging the LLMs' strong performance in interpreting and responding to prompts. For instance, BLIP-2~\cite{Li2023BLIP2BL} introduces the Q-Former to synchronize vision foundation models with LLMs without modifying the underlying models. MiniGPT4~\cite{Zhu2023MiniGPT4EV} utilizes a straightforward fully connected layer, requiring only a minimal set of caption data. LLaVA~\cite{Liu2023VisualIT}  enhances the LLM with high-quality instructional data generated by GPT-4. QWen-VL~\cite{Qwen-VL} undergoes fine-tuning with high-resolution images, employing multi-task training strategies. Moreover, mPLUG-DocOwl~\cite{Ye2023mPLUGOwlME} expands the capabilities of LVLMs to include document understanding by processing digital document data. Given that instructions for LVLMs challenge both the vision and language processing capabilities simultaneously, there is a pressing need for a benchmark designed to evaluate the two functions in a nuanced manner at the same time. ConvBench would facilitate a comprehensive assessment of LVLMs' abilities in tasks that demand intricate coordination between visual perception and linguistic interpretation.

\subsubsection{Large Vision-Language Models Benchmarks.} 
With the advancement of Vision-Language Models (LVLMs), existing standard evaluation benchmarks like MSCOCO~\cite{Lin2014MicrosoftCC}, GQA~\cite{Hudson2019GQAAN}, VQA~\cite{Agrawal2015VQAVQ, Goyal2016MakingTV},  etc., are no longer sufficient to assess the comprehensive multimodal abilities of LVLMs. In response, a variety of benchmarks have been developed specifically for LVLM evaluation, including OwlEval~\cite{Ye2023mPLUGOwlME}, LAMM~\cite{Yin2023LAMMLM}, LVLM-eHub~\cite{Xu2023LVLMeHubAC}, SEED~\cite{Li2023SEEDBenchBM}, MMBench~\cite{Liu2023MMBenchIY}, and MM-Vet~\cite{Yu2023MMVetEL}. These benchmarks primarily focus on assessing basic perceptual abilities. In addition, VisIT-Bench~\cite{Bitton2023VisITBenchAB} covers a broad spectrum of tasks, ranging from simple recognition to complex reasoning. Recent research has also introduced LVLM benchmarks requiring expert-level domain knowledge and intricate reasoning, such as MathVista~\cite{Lu2023MathVistaEM}, MMMU~\cite{Yue2023MMMUAM} and MMT-Bench~\cite{ying2024mmtbench}. However, these benchmarks tend to address perception, reasoning, and creation tasks in isolation, without establishing connections among these tasks. Furthermore, the current benchmarks predominantly focus on single-turn interactions, comprising one instruction and one response, with less emphasis on multi-turn interactions between users and chatbots. The ConvBench addresses these gaps by not only offering a progressive evaluation that moves from basic perception through logical reasoning to advanced creation but also by evaluating LVLMs' capabilities in multi-turn conversational contexts.

\begin{figure}[t]
  \centering
  \includegraphics[width=\textwidth]{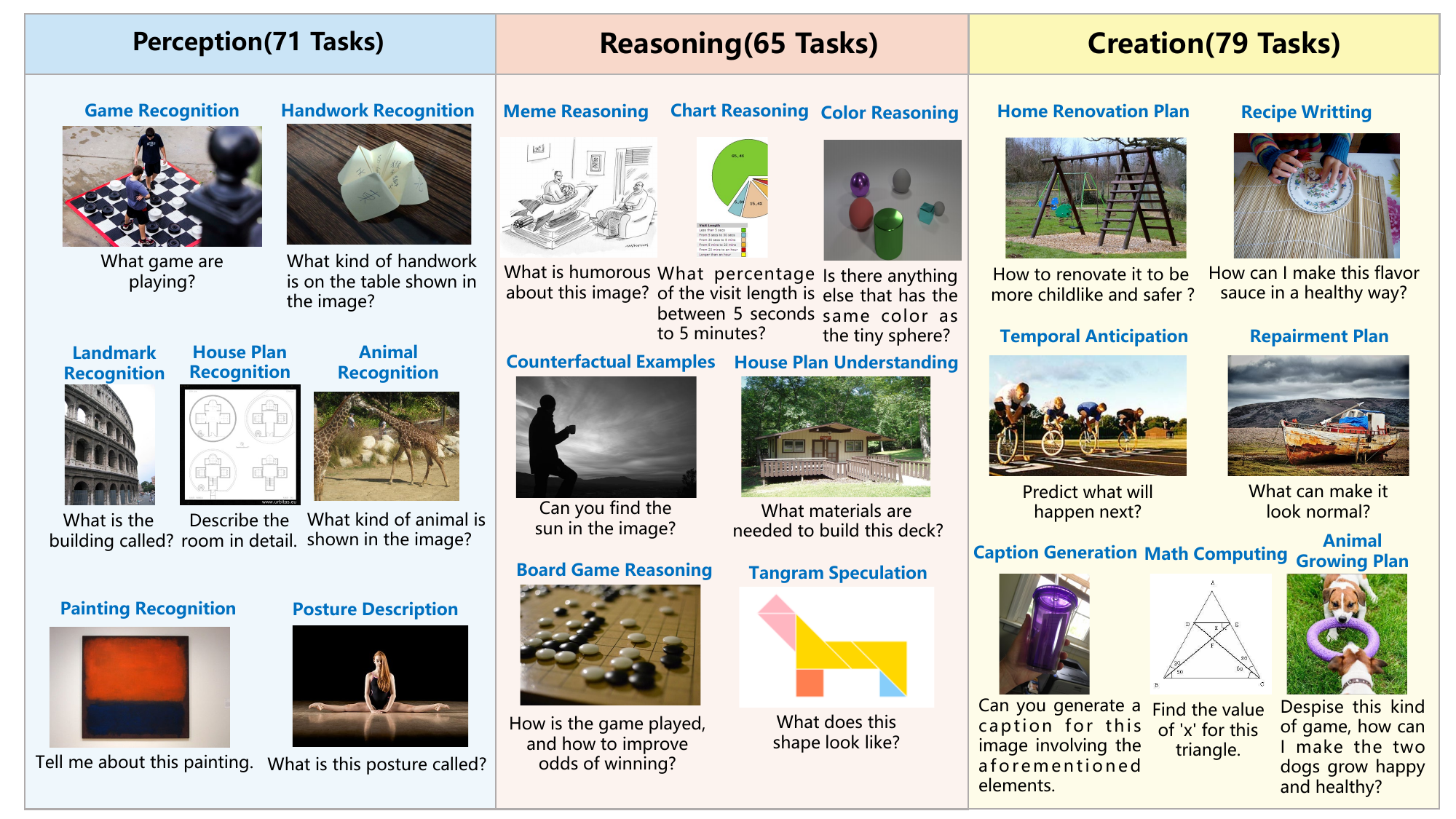} 
  \caption{Visualization of example tasks in ConvBench. It consists of 215 tasks constructed in perception, reasoning, and creation hierarchy.}
  \label{fig:tasks}
\end{figure}

\section{ConvBench}
\subsection{Overview of ConvBench}
We introduce the ConvBench Benchmark, an innovative benchmark designed to evaluate multi-turn conversation capabilities and conduct detailed, progressive assessments of Large Vision-Language Models (LVLMs). This benchmark concentrates on three pivotal skills of LVLMs: perception, reasoning, and creation. These skills are meticulously arranged in a bottom-up way, facilitating error attribution. 

The ConvBench includes 577 premium image-instruction pairs tailored for multi-turn dialogues. Each pair is structured around three sequential instructions, each targeting a distinct cognitive skill—beginning with perception, followed by reasoning, and culminating in creation. This structure underscores the cognitive evolution from basic perceptual comprehension to logical reasoning and finally to sophisticated creative expression. Encompassing 215 tasks, the benchmark is divided into 71 tasks focused on perception, 65 on reasoning, and 79 on creation. Drawn from real-world contexts, these demanding tasks are seen as benchmarks for the competencies of advanced LVLMs, as depicted in Figure~\ref{fig:tasks}. Beyond the instructions, we manually annotate each instruction's referenced answer, which facilitates error attribution and automatic evaluation in our evaluation framework, ConvEval.


\begin{figure}[t]
  \centering
  \includegraphics[width=1.0\textwidth]{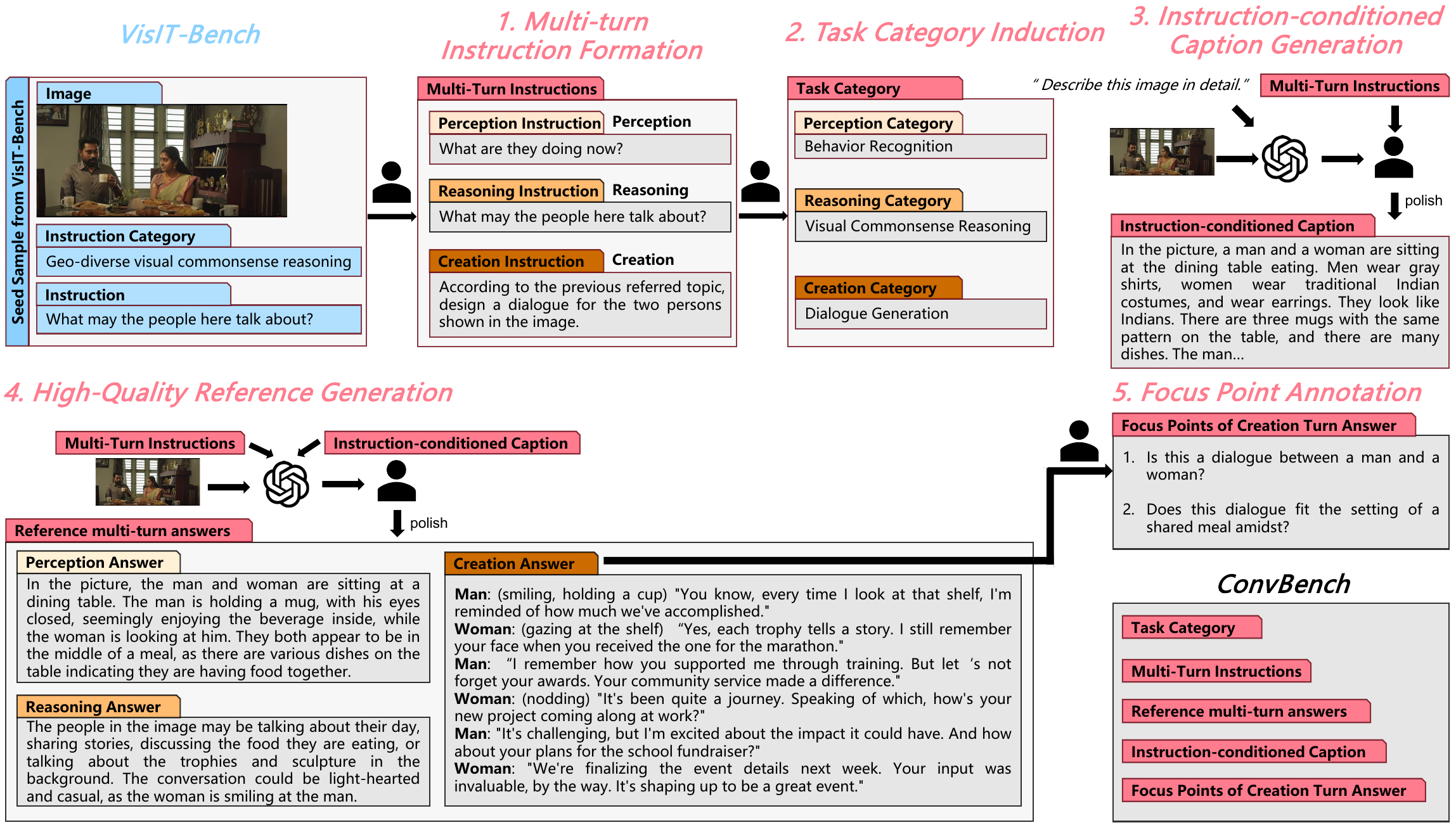} 
  \caption{The pipeline of data curation. We develop three multi-turn instructions for each image to assess the perception, reasoning, and creation capabilities. We also annotate the referenced answers to facilitate automatic evaluation and error analysis.}
  \label{fig:data_curation_process}
\end{figure}

\subsection{Data Curation Process}
\textbf{Data Collection.} Our benchmark collection is structured into five distinct stages, as depicted in Figure~\ref{fig:data_curation_process}. These stages are as follows: 

\textbf{i) Multi-turn Instruction Formation.} Starting with VisIT-Bench as our foundational guide, we develop three-level instructions for each sample from VisIT-Bench. The three-level instructions align with perception, reasoning, and creation capability. 

\textbf{ii) Task Category Induction.} We derive the task categories by inducing them from instructions. The bottom-up approach to task collection guarantees that the tasks under investigation are tailored to meet real-world requirements. ConvBench consists of 215 tasks, and we have included the details in the supplymentarl materials.

\textbf{iii) Instruction-Conditioned Caption Annotation.} We then proceed to annotate detailed captions for images tailored to the instructions. We first prompt GPT-4V with "Describe this image in detail." We then polish the responses according to the instructions to obtain the final instruction-conditioned caption. These captions serve a dual purpose: they provide a comprehensive description of the image relevant to executing the instructions and support the generation of raw reference answers in the next step. Additionally, these annotated captions are instrumental in evaluating responses from candidate models, ensuring a robust assessment framework. 

\textbf{iv) High-Quality Reference Generation.} For each sample, we feed GPT-4V with the instruction-conditioned caption, the image, multi-turn instructions, and our well-designed prompt in a multi-turn conversation fashion to generate each instruction's response.
We meticulously refine these responses as reference answers, enhancing their quality and relevance. 

\textbf{v) Focus Point Annotation.} The creativity instruction is an open-ended question without a standard answer. Therefore, we annotate specific focus points related to each creation instruction. These annotations are used as criteria to assess whether the model produces instructive answers to the instruction, seeing Step 5 in Figure~\ref{fig:data_curation_process}. 

\begin{figure}[t]
  \centering
  \includegraphics[width=\textwidth]{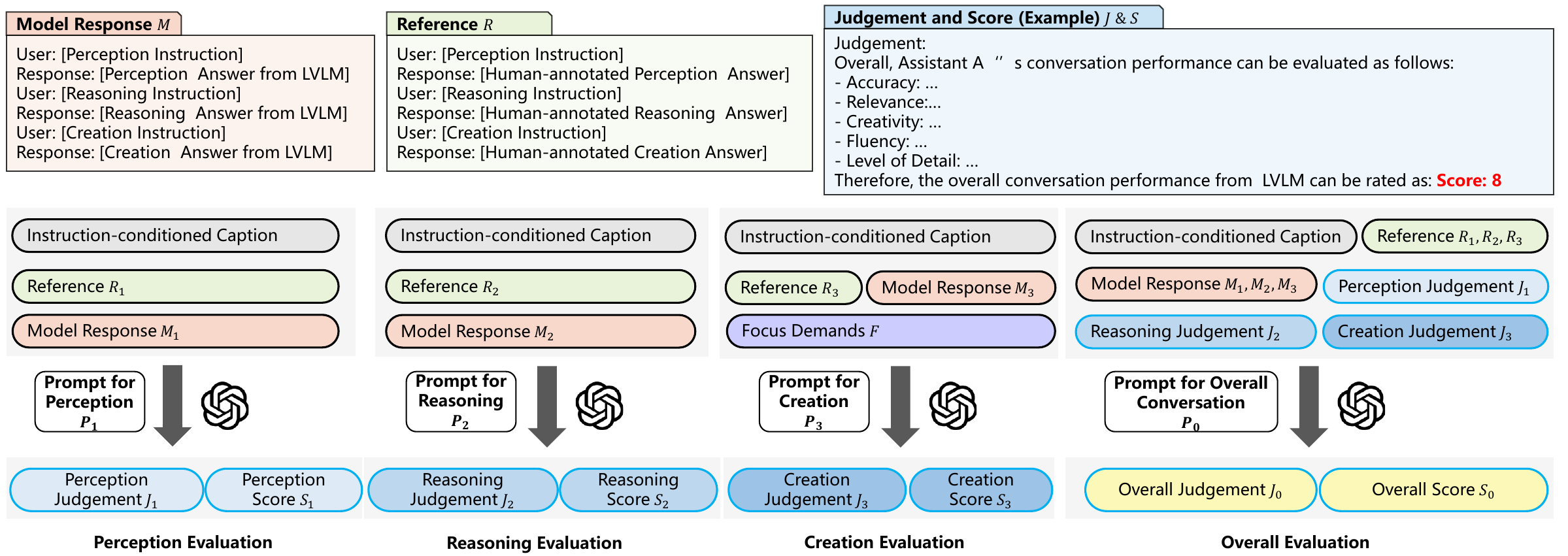} 
  \caption{ConvBench Evaluation Pipeline. \includegraphics[height=0.75em]{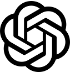} indicates the ChatGPT-3.5.} 
  \label{fig:evaluation_pipeline}
\end{figure}

\section{Multi-turn Conversation Evaluation}

In this section, we introduce ConvEval, an evaluation pipeline designed for multi-turn conversation assessment. Building upon prior research~\cite{Bitton2023VisITBenchAB, Zheng2023JudgingLW}, which has shown the effectiveness of employing Large Language Models (LLMs) with tailored prompts for automatic evaluation, ConvEval utilizes a large language model (LLM) as the evaluator to assess model responses across various levels, ensuring cost-effectiveness and efficiency in evaluation procedures.

Specifically, ConvEval comprises two kinds of grading schemes for models' responses: direct and pairwise grading. The direct grading will give a score of $0$-$10$ for the model's responses under specific prompts while the pairwise grading would output a model preference when comparing the response of the test model and reference. In the sequel, we introduce the evaluation pipeline in Sec. \ref{sec:conveval-pipeline} and present direct and pairwise grading in Sec. \ref{sec:conveval-pair} and Sec. \ref{sec:conveval-direct}, respectively.


\subsection{Pipeline of ConvEval} \label{sec:conveval-pipeline}

ConEval comprises four key components: perception, reasoning, creation, and overall conversation evaluation modules, as shown in Figure \ref{fig:evaluation_pipeline}. The multi-turn conversation evaluation is complicated as creativity may be affected by inaccurate perception or reasoning in previous levels. 
To enable error attribution, we recursively use ConvEval in three settings as follows. For clarity, we denote the instruction, model response, reference, and focal points as $I_i, M_i, R_i,$ and $F_i$ at each level, respectively, where $i$ indicates the level index. Note that $F_1$ and $F_2$ are null focal points.

\textbf{i) ConvEval with model's responses.} In this setting, the model response at each level is obtained by $M_i = f(I_0^{i-1}, M_0^{i-1}, I_i), i=1,2,3$ where $f$ denotes the model inference function. The evaluation process can be expressed as $S_i, J_i= \mathrm{ConvEval}(\{M_i\}_{i=1}^3, \{R_i\}_{i=1}^3, \{F_i\}_{i=1}^3; P_i), i=1,2,3$ where $S_i, J_i,$ and $P_i$ are the overall capability score, judgment from the LLM, and prompt specific to level $i$. Finally, we feed all instructions, responses, focal points, and judgments into LLM to obtain the overall conversation score as given by 
$S_0, J_0 =\mathrm{ConvEval}(\{M_i\}_{i=1}^3, \{R_i\}_{i=1}^3, \{J_i\}_{i=1}^3; P_0)$ where $P_0$ is the prompt to obtain $S_0$.

\textbf{ii) ConvEval with perfect perception reference.} To enable error attribution, we first use perfect perception reference as the response at the perception level. In this way, the influence of inaccurate perception on reasoning, creation, and overall conversation can be derived. In this setting, the model inference can be written as $\hat{M}_1 = R_1$ and $\hat{M}_i = f(I_0^{i-1}, \hat{M}_0^{i-1}, I_i), i=2,3$. The evaluation process without considering perception error can be expressed as $\hat{S}_i, \hat{J}_i= \mathrm{ConvEval}(\{\hat{M}_i\}_{i=1}^3, \{R_i\}_{i=1}^3, \{F_i\}_{i=1}^3; P_i), i=2,3$. Finally, the overall conversation score without considering perception score can be given by $\hat{S}_0, \hat{J}_0 =\mathrm{ConvEval}(\{\hat{M}_i\}_{i=1}^3, \{R_i\}_{i=1}^3, \{\hat{J}_i\}_{i=2}^3; P_0)$. By comparing ${S}_i$ and $\hat{S}_i$ ($i=0,2,3)$, we can see how perception error affects the performance of overall conversation, reasoning and creativity.

\textbf{iii) ConvEval with perfect perception and reasoning reference.} We further explore how reasoning errors affect creativity and overall conversation. To this end, we use perfect perception and reasoning reference as responses at perception and reasoning levels. In this setting, the model inference can be written as $\tilde{M}_i = R_i, i=1,2$ and $\tilde{M}_i = f(I_0^{i-1}, \tilde{M}_0^{i-1}, I_i), i=3$. The evaluation process without considering perception and reasoning error can be expressed as $\tilde{S}_i, \tilde{J}_i= \mathrm{ConvEval}(\{\tilde{M}_i\}_{i=1}^3, \{R_i\}_{i=1}^3, \{F_i\}_{i=1}^3; P_i), i=3$. Finally, the overall conversation score without considering perception and reasoning score can be given by $\tilde{S}_0, \tilde{J}_0 =\mathrm{ConvEval}(\{\tilde{M}_i\}_{i=1}^3, \{R_i\}_{i=1}^3, \tilde{J}_3; P_0)$. By comparing $\hat{S}_i$ and $\tilde{S}_i$ ($i=0,3)$, we can check how reasoning errors affect the performance of overall conversation and creativity.

Note that the function in i) - iii) $\mathrm{ConvEval}(\cdot)$ can be pairwise or direct grading schemes as described in the following.

\subsection{Pairwise Grading}\label{sec:conveval-pair}
When $\mathrm{ConvEval}(\cdot)$  employs pairwise grading scheme,  a model preference is returned when comparing the response of the test model and reference. Following the evaluation pipeline in Sec. \ref{sec:conveval-pipeline}, we feed LLM with the following components for evaluation: system prompts, instructions, model responses, human-verified references, and focal points at all levels. To make sure that LLM can infer the correct answers, the instruction-condition caption tailored to the specific instructions is also added.


The response sets are anonymously presented to the LLM in a random order. The LLM is tasked with a pairwise comparison to decide which set of responses is superior. The percentages of cases where the LLM prefers the output from the model rather than the human-verified reference output are obtained as the final metric, \emph{i.e.} win rate. The system prompts, detailed in Appendix, encourage the LLM to engage in a step-by-step thought process, making its reasoning explicit. In our forced-choice setup, ties are not permitted; thus, if the LLM deems the responses of equal quality, it is instructed to select one arbitrarily. This prompt, including the presentation of two full conversations within a single prompt, addresses the challenge of the LLM potentially struggling to accurately recall previous responses by the assistant, as noted in previous work~\cite{Zheng2023JudgingLW}.

\subsection{Direct Grading}\label{sec:conveval-direct}

When $\mathrm{ConvEval}(\cdot)$ employs the pairwise grading scheme,  a score of $0$-$10$ is returned when comparing the response of the test model and reference. Unlike pairwise comparison, the response sets are now identifiable; specifically, the sets from the tested LVLM and a human participant are labelled as Assistant A's Conversation and Reference Answer, respectively. In this context, the LLM judge is tasked with assigning scores directly to each turn answer and the overall conversation quality.

The prompt designed for these scenarios is detailed in Appendix. This prompt not only encourages the LLM to engage in a chain-of-thought process but also includes the presentation of two full conversations within a single prompt. This approach is aimed at enhancing the LLM's ability to accurately evaluate and score the responses by maintaining a clear context and facilitating a comprehensive assessment.

\section{Experiment and Analysis}
We undertake a thorough evaluation of $19$ LVLMs using the ConvBench. Section 5.1 outlines the evaluation framework, detailing the LVLMs under study and the methodologies employed for assessment. Section 5.2 delves into progressive evaluation comparisons and analysis, offering insights into how these models perform over a range of tasks. Section 5.3 focuses on multi-turn conversation comparisons and analysis, examining the models' capabilities in engaging in dialogues that require sustained interaction.

\subsection{Evaluation Settings}
\textbf{LVLMs.} We evaluate $19$ representative LVLMs, including GPT-4V~\cite{GPT4V}, GeminiProVision~\cite{Gemini}, Reka Flash~\cite{Reka}, Qwen-VL-Chat~\cite{Qwen-VL}, LLaMA-Adapter-v2~\cite{Gao2023LLaMAAdapterVP}, XComposer~\cite{Zhang2023InternLMXComposerAV}, XComposer2~\cite{Dong2024InternLMXComposer2MF},  mPLUG-Owl2~\cite{Ye2023mPLUGOwlME}, Monkey~\cite{Li2023MonkeyIR}, Otter~\cite{Li2023OtterAM},  MMAlaya~\cite{datacanvas2024mmalaya}, MiniGPT-4~\cite{Zhu2023MiniGPT4EV}, InternVL-Chat-V1-2~\cite{Chen2023InternVLSU}, InternVL-Chat-V1-5~\cite{Chen2023InternVLSU}, LLaVA (V1.5) models~\cite{Liu2023VisualIT} like LLaVA-7B, LLaVA-13B, ShareGPT4V models~\cite{Chen2023ShareGPT4VIL} like ShareGPT4V-7B, ShareGPT4V-13B, BLIP2 models~\cite{Li2023BLIP2BL} like BLIP2-FLAN-T5-XL and BLIP2-FLAN-T5-XXL. Appendix provides the models' configuration.

\textbf{Evaluation Methods.} We supplied each Large Vision-Language Model (LVLM) with an image accompanied by a set of carefully curated progressive instructions to elicit corresponding responses. The chatting prompt specific to each model played a crucial role in the generation of multi-turn responses. Upon gathering all responses, we employed our proposed $ConvBenchEval$ methodology to perform a quantitative analysis. This involved comparing the model-generated responses with high-quality, human-verified reference answers. We utilized $9$ distinct evaluation scores as metrics to assess the multi-turn conversation capabilities and to provide detailed, progressive evaluation results.

\begin{table*}[t!]
\centering 
    \caption{Comparison of Performance for LVLMs on ConvBench. Quantitative ConvBench Evaluation Results for $19$ LVLMs with Pairwise Grading method. The results in the table are win-rate vs human. The colors blue and red indicate positive and negative differences, respectively. $R_{2}$ is defined as $(S_{1}+S_{2}+S_{3})/{3}$, indicative of the mean performance over three turns. Meanwhile, $R_{1}$ is computed as $(R_{2}+S_{0})/{2}$, representing the model's overall score.}
    \label{tab:pairwise_results}
\resizebox{1.0\textwidth}{!}{
\begin{tabular}{l|ll|lllllllll}
\toprule
{Model} & $R_{1}$ & $R_{2}$ & $S_{1}$ &  $S_{2}$ & $S_{3}$ & $S_0$ & $\hat{S}_{2}$$(\hat{S}_{2}-S_{2})$ & $\hat{S}_{3}$$(\hat{S}_{3}-S_{3})$ & $\hat{S}_{0}$$(\hat{S}_{0}-S_0)$ & $\tilde{S}_{3}$$(\tilde{S}_{3}-\hat{S}_{3})$ & $\tilde{S}_{0}$$(\tilde{S}_{0}-\hat{S}_{0})$ \\
 \midrule
 \midrule
{GPT-4V} & \textbf{39.51} & \textbf{38.47}  & \textbf{38.47} & \textbf{39.34} & \textbf{37.61} & \textbf{40.55} & \textbf{47.31}(\blue{16.97}) & 37.78(\blue{0.61})  & 37.61(\red{2.94})  & 38.99(\blue{1.21}) & 38.30(\blue{0.69})    \\\midrule
{Claude} & 36.60 & 37.49  & \textbf{38.99} & 39.17 & 34.32 & 35.70 & 45.93(\blue{6.76}) & \textbf{38.99}(\blue{4.67})  & \textbf{43.15}(\blue{7.45})  & \textbf{39.16}(\blue{0.17}) & \textbf{40.21}(\red{2.94})    \\\midrule
{Reka Flash} & 25.60 & 24.67  & 25.13 & 27.56 & 21.32 & 26.52 & 32.93(\blue{5.37}) & 22.88(\blue{1.56}) & 25.82(\red{0.70}) & 24.78(\blue{1.90}) & 26.00(\blue{0.18}) \\\midrule
{InternVL-Chat-V1-2 } & 21.17 & 22.41  & 24.96 & 21.31 & 20.97 & 19.93 & 32.06(\blue{10.75})  & 28.25(\blue{7.28})  & 29.64(\blue{9.71})  & 33.62(\blue{5.37}) & 35.18\blue{(5.54})  \\\midrule
{ShareGPT4V-13B } & 17.56 & 17.45  & 17.85 & 18.72 & 15.77 & 17.68 & 32.58(\blue{13.86})  & 30.33(\blue{14.56})  & 28.94(\blue{11.26})  & 32.41(\blue{2.08}) & 31.54(\blue{2.60}) \\\midrule
{LLaVA-V1.5-13B } & 16.93 & 18.08 & 20.45 & 18.02 & 15.77 & 15.77 & 32.76(\blue{14.74}) & 25.65(\blue{9.88})  & 28.94(\blue{13.17})  & 32.06(\blue{6.41}) & 28.94(\blue{0.00})\\\midrule
{ShareGPT4V-7B } & 16.32 & 16.87  & 16.81 & 19.24 & 14.56 & 15.77 & 32.76(\blue{13.52}) & 23.05(\blue{8.49})  & 25.13(\blue{9.36})  & 29.46(\blue{6.41}) & 30.33(\blue{5.20}) \\\midrule
{LLaVA-V1.5-7B } & 16.15 & 17.56  & 19.06 & 19.24 & 14.38 & 14.73 & 33.80(\blue{14.56})  & 23.22(\blue{8.84})  & 26.52(\blue{11.79})  & 30.68(\blue{7.46}) & 32.58(\blue{6.06})  \\\midrule
{XComposer2} & 15.83 & 16.41  & 17.16 & 19.06 & 13.00 & 15.25 & 30.50(\blue{11.44}) & 20.97(\blue{7.97})  & 22.36(\blue{7.11})  & 28.60(\blue{7.63}) & 29.81(\blue{7.45}) \\\midrule
{mPLUG-Owl2} & 14.93 & 15.83  &  17.50 & 17.16 & 12.82 & 14.04 & 27.90(\blue{10.74}) & 17.50(\blue{4.68})  & 20.80(\blue{6.76})  & 24.26(\blue{6.76}) & 24.44(\blue{3.64}) \\\midrule
{Qwen-VL-Chat} & 14.33 & 14.62  & 16.29 & 18.37 & 9.19 & 14.04 & 28.25(\blue{9.88})  & 16.12(\blue{6.93})  & 22.70(\blue{8.66})  & 25.30(\blue{9.18}) & 26.52(\blue{3.82}) \\\midrule
{MiniGPT-4} & 10.95 & 10.80 &  11.61 & 11.27 & 9.53 & 11.09 & 27.56(\blue{16.29}) & 18.20(\blue{8.67})  & 22.53(\blue{11.44})  & 22.88(\blue{4.68}) & 23.74(\blue{1.21}) \\\midrule
{LLaMA-Adapter-v2} & 9.04 & 9.59  & 8.84 & 10.92 & 9.01 & 8.49 & 27.38(\blue{16.46})  & 15.60(\blue{6.59})  & 19.41(\blue{10.92})  & 18.37(\blue{2.77}) & 19.24(\red{0.17}) \\\midrule
{GeminiProVision } & 8.44 & 8.55  & 9.01 & 9.36 & 7.28 & 8.32 & 21.84(\blue{12.48}) & 12.31(\blue{5.03})  & 15.08(\blue{6.76})  & 23.92(\blue{11.61}) & 23.92(\blue{8.84})\\\midrule
{MMAlaya} & 5.55 & 5.89 & 7.28 & 6.41 & 3.99 & 5.20 & 22.53(\blue{16.12}) & 9.88(\blue{5.99})  & 15.25(\blue{10.05})  & 14.21(\blue{4.33}) & 16.81(\blue{1.56})    \\\midrule
{Monkey} & 3.70 & 4.10 & 3.64 & 5.20 & 3.47 & 3.29 & 16.64(\blue{11.44}) & 7.28(\blue{3.81})  & 10.75(\blue{7.46})  & 13.86(\blue{6.58}) & 15.94(\blue{5.19})  \\\midrule
{Otter} &  2.78 & 2.60 & 3.12 & 3.12 &  1.56 & 2.95 & 14.21(\blue{11.09}) & 5.37(\blue{3.81})  & 9.01(\blue{6.06})  & 8.49(\blue{3.12}) & 13.00(\blue{3.99}) \\\midrule
{XComposer} & 1.21 & 1.73  & 1.73 & 1.91 & 1.56 & 0.69 & 12.13(\blue{10.22}) & 2.77(\blue{1.21})  & 8.49(\blue{7.80})  & 10.40(\blue{7.63}) & 12.48(\blue{3.99}) \\\midrule
{BLIP2-FLAN-T5-XXL } & 0.32 & 0.29  & 0.35 & 0.52 & 0.00 & 0.35 & 3.47(\blue{2.95})  & 1.91(\blue{1.91})  & 2.95(\blue{2.60})  & 5.72(\blue{3.81}) & 8.49(\blue{5.54})\\\midrule
{BLIP2-FLAN-T5-XL } & 0.06 & 0.11  & 0.00 & 0.17 & 0.17 & 0.00 & 3.12(\blue{2.95})  & 0.17(\blue{0.00})  & 2.25(\blue{2.25})  & 3.97(\blue{3.80}) & 8.67(\blue{6.42}) \\
 \bottomrule
 \end{tabular}}
 \end{table*}

\begin{table*}[t!]
\centering 
\caption{
Comparison of Performance for LVLMs on ConvBench. Quantitative ConvBench Evaluation Results for $19$ LVLMs with Direct Grading method. The results in the table is the average scores of all the samples. The colors blue and red indicate positive and negative differences, respectively. $R_{2}$ is defined as $(S_{1}+S_{2}+S_{3})/{3}$, indicative of the mean performance over three turns. Meanwhile, $R_{1}$ is computed as $(R_{2}+S_{0})/{2}$, representing the model's overall score.}
\label{tab:single_results}
\resizebox{1.0\textwidth}{!}{%
\begin{tabular}{l|ll|lllllllll}
\toprule
{Model} & $R_{1}$ & $R_{2}$ & $S_{1}$ &  $S_{2}$ & $S_{3}$ & $S_0$ & $\hat{S}_{2}$$(\hat{S}_{2}-S_{2})$ & $\hat{S}_{3}$$(\hat{S}_{3}-S_{3})$ & $\hat{S}_{0}$$(\hat{S}_{0}-S_0)$ & $\tilde{S}_{3}$$(\tilde{S}_{3}-\hat{S}_{3})$ & $\tilde{S}_{0}$$(\tilde{S}_{0}-\hat{S}_{0})$ \\
 \midrule
 \midrule
{GPT-4V} & \textbf{7.09} & \textbf{7.30}  & \textbf{7.30} & \textbf{7.48} & \textbf{7.12} & \textbf{6.88} & \textbf{8.23}(\blue{0.75}) & \textbf{8.00}(\blue{0.88}) & \textbf{8.25}(\blue{1.37}) & \textbf{7.34}(\red{0.66}) & \textbf{8.18}(\red{0.07})    \\\midrule
{Claude} & 6.54 & 6.75 & 6.53 & 7.04 & 6.68 & 6.32 & 7.48(\blue{0.44}) & 7.06(\blue{0.38}) & 7.55(\blue{1.23}) & 7.18(\blue{0.12}) & 8.13(\blue{0.58}) \\\midrule
{Reka Flash} & 6.78 & 6.86 & 6.93 & 7.25 & 6.41 & 6.70 & 7.10(\red{0.15}) & 6.41(\blue{0.00}) & 7.32(\blue{0.62}) & 4.95(\red{1.46}) & 6.95(\red{0.37}) \\\midrule
{ShareGPT4V-7B } & 5.83 & 5.99  & 6.02 & 6.14 & 5.80 & 5.67 & 7.19(\blue{1.05}) & 6.77(\blue{0.97}) & 7.31(\blue{1.64}) & 6.93(\blue{0.16}) & 8.19(\blue{0.88}) \\\midrule
{XComposer2} & 5.82 & 5.98  & 5.98 & 6.17 & 5.78 & 5.66 & 7.35(\blue{1.18}) & 7.04(\blue{1.26}) & 7.66(\blue{2.00}) & 7.00(\red{0.04}) & 8.20(\blue{0.54}) \\\midrule
{Qwen-VL-Chat} & 5.54 & 5.65  & 5.96 & 5.78 & 5.22 & 5.43 & 7.04(\blue{1.26}) & 6.53(\blue{1.31}) & 7.26(\blue{1.83}) & 6.57(\blue{0.04}) & 8.00(\blue{0.74}) \\\midrule
{InternVL-Chat-V1-2 } & 5.49 & 5.69  & 5.80 & 5.88 & 5.39 & 5.29 & 6.66(\blue{0.78}) & 6.12(\blue{0.73}) & 6.75(\blue{1.46}) & 6.31(\blue{0.19}) & 7.70(\blue{0.95}) \\\midrule
{LLaVA-V1.5-7B } & 5.16 & 5.29  & 4.95 & 5.59 & 5.34 & 5.03 & 7.28(\blue{1.69}) & 6.68(\blue{1.34}) & 7.28(\blue{2.25}) & 6.72(\blue{0.04}) & 7.97(\blue{0.69}) \\\midrule
{mPLUG-Owl2} & 5.04 & 5.17  & 4.98 & 5.38 & 5.14 & 4.91 & 6.77(\blue{1.39}) & 6.64(\blue{1.50}) & 7.22(\blue{2.31}) & 5.93(\red{0.71}) & 7.62(\blue{0.40}) \\\midrule
{LLaVA-V1.5-13B } & 4.94 & 5.14 & 5.03 & 5.41 & 4.99 & 4.74 & 7.43(\blue{2.02}) & 7.13(\blue{2.14}) & 7.70(\blue{2.95}) & 6.14(\red{0.99}) & 7.60(\red{0.10}) \\\midrule
{ShareGPT4V-13B } & 4.85 & 5.03  & 5.16 & 5.06 & 4.86 & 4.67 & 7.42(\blue{2.36}) & 7.17(\blue{2.31}) & 7.65(\blue{2.98}) & 6.24(\red{0.93}) & 7.65(\blue{0.00}) \\\midrule
{LLaMA-Adapter-v2}& 4.77 & 4.91 & 4.77 & 5.47 & 4.48 & 4.64 & 6.68(\blue{1.21}) & 5.49(\blue{1.01}) & 6.68(\blue{2.04}) & 5.19(\red{0.30}) & 7.36(\blue{0.68}) \\\midrule
{Monkey} & 4.49 & 4.60 & 5.11 & 4.68 & 4.01 & 4.37 & 6.28(\blue{1.60}) & 5.66(\blue{1.65}) & 6.76(\blue{2.39}) & 5.39(\red{0.27}) & 7.30(\blue{0.54})   \\\midrule
{GeminiProVision } & 4.42 & 4.60  & 5.18 & 4.95 & 3.66 & 4.24 & 6.16(\blue{1.21}) & 5.05(\blue{1.39}) & 6.28(\blue{2.04}) & 5.07(\blue{0.02}) & 7.05(\blue{0.77}) \\\midrule
{MiniGPT-4} & 3.85 & 4.04  & 3.99 & 4.40 & 3.73 & 3.66 & 6.66(\blue{2.26}) & 5.80(\blue{2.07}) & 6.75(\blue{3.09}) & 4.97(\red{0.83}) & 7.01(\blue{0.26})    \\\midrule
{MMAlaya} & 3.60 & 3.75  & 4.07 & 3.91 & 3.28 & 3.44 & 5.64(\blue{1.73}) & 4.76(\blue{1.48}) & 5.91(\blue{2.47}) & 4.02(\red{0.74}) & 6.47(\blue{0.56})    \\\midrule
{Otter} & 2.96 & 3.11 & 3.33 & 3.52 &  2.47 & 2.80 & 5.00(\blue{1.48}) & 4.11(\blue{1.64}) & 5.75(\blue{2.95}) & 3.25(\red{0.86}) & 6.03(\blue{0.28})  \\\midrule
{XComposer} & 2.61 & 2.70  & 2.90 & 2.82 & 2.39 & 2.51 & 4.67(\blue{1.85}) & 3.84(\blue{1.45}) & 5.30(\blue{2.79}) & 3.90(\blue{0.06}) & 6.47(\blue{1.17})  \\\midrule
{BLIP2-FLAN-T5-XXL } & 2.37 & 2.45  & 2.81 & 2.59 & 1.95 & 2.28 & 3.18(\blue{0.59}) & 2.35(\blue{0.40}) & 4.03(\blue{1.75}) & 2.59(\blue{0.24}) & 5.75(\blue{1.72}) \\\midrule
{BLIP2-FLAN-T5-XL } & 2.14 & 2.21  & 2.55 & 2.33 & 1.74 & 2.07 & 2.74(\blue{0.41}) & 2.17(\blue{0.43}) & 3.81(\blue{1.74}) & 2.24(\blue{0.07}) & 5.44(\blue{1.63}) \\

 \bottomrule
\end{tabular}%
}
\end{table*}

\subsection{Progressive Evaluation Comparisons and Analysis}
The outcomes of the evaluation results are detailed in Table~\ref{tab:pairwise_results} and Table~\ref{tab:single_results}. In these tables, $S_{1}$, $S_{2}$, and $S_{3}$ denote the scores for perception, reasoning, and creation, respectively. Meanwhile, $\hat{S}_{2}$ and $\hat{S}_{3}$ correspond to the scores for reasoning and creation, respectively, but under conditions of perfect perception. $\tilde{S}_{3}$ is the score for creation, assuming perfect conditions for both perception and reasoning. We present our principal insights from the evaluation results as follows:

(1) \textbf{The Challenge of Progressive Evaluation:} This benchmark sets formidable challenges for modern models. GPT-4V, despite being a sophisticated model, shows only modest achievements in perception, reasoning, and creation. In the Pairwise Grading approach, it attains scores of 38.47, 39.34, and 37.61 for perception, reasoning, and creation, respectively. On the other hand, through the Direct Grading method, it achieves 7.30, 7.48, and 7.12, correspondingly. Claude~\cite{Claude} is infinitely close to GPT-4V in performance. However, according to progressive evaluation, Claude~\cite{Claude} has greater potential for performance improvement compared to GPT-4V. The least successful models, exemplified by BLIP2, are utterly unsuccessful, underscoring a significant opportunity for advancements. Our analysis exposes a stark discrepancy between the performance of these models and that of humans, highlighting the benchmark's stringent and exacting criteria.

\begin{figure}[t]
  \centering
  \includegraphics[width=1.0\textwidth]{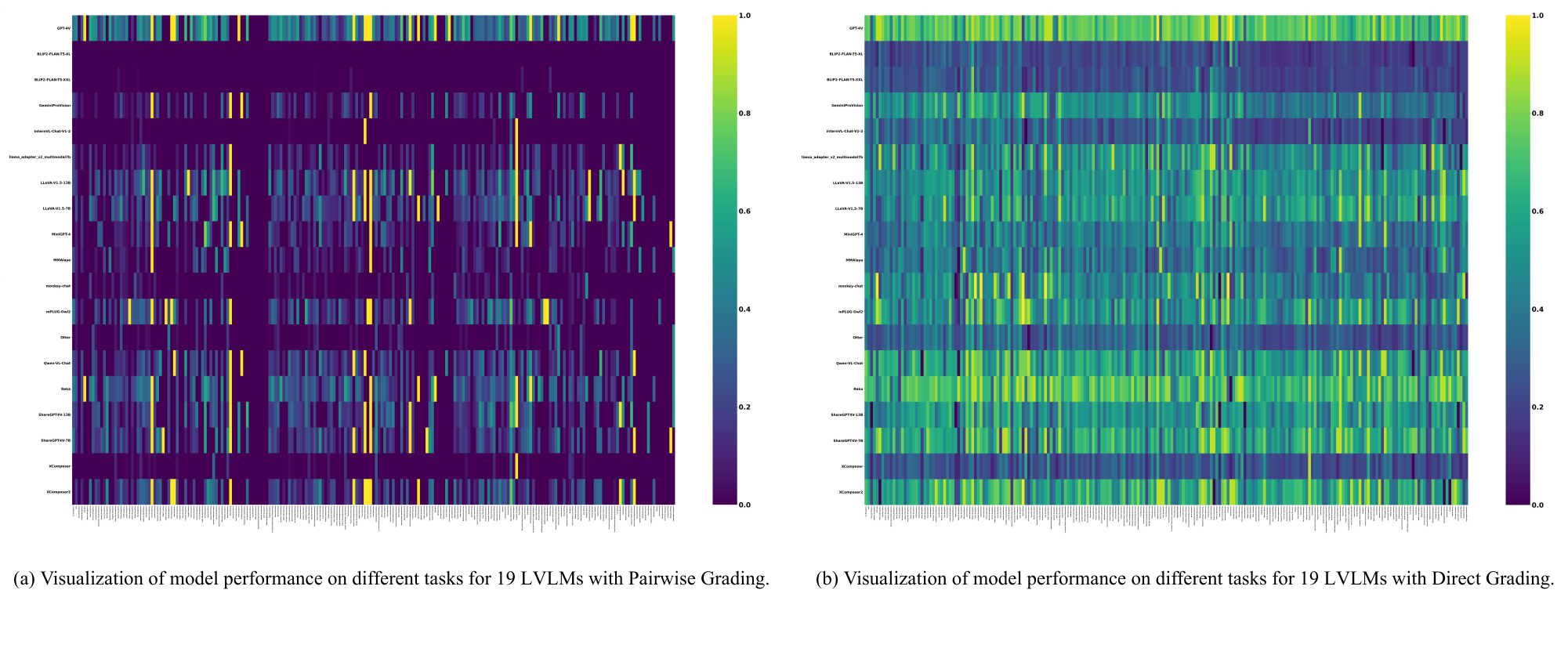} 
  \caption{The Visualizations of model performance. The horizontal and vertical axes represent task and model names respectively. The performance heatmaps across various tasks for $19$ LVLMs, under the two grading strategies. They have high agreement in comparing the performance of different models. GPT-4V and Reka have excellent performance. BLIP2, Otter and Xcomposer have poor performance. However, Figure(a) is darker and has a lower score than Figure(b), which may be due to the Direct Grading method struggles to identify subtle differences between specific pairs resulting in higher scores.}
\label{fig:heatmaps}
\end{figure}

(2) \textbf{Weak Perception Undermines LVLMs' Reasoning and Creation Performance:}
Under conditions of perfect perception, we see significant improvements in reasoning and creation abilities, as indicated by the data in the $\hat{S}_{2}$ and $\hat{S}_{3}$ columns of Table~\ref{tab:pairwise_results} and Table~\ref{tab:single_results}. The figures in parentheses reflect the enhancement in reasoning and creation attributed to impeccable perception. Across $19$ LVLMs, the average increase in reasoning and creation scores are 11.21 and 5.31, respectively, when using the Pairwise Grading method. Similarly, improvements of 1.25 both are observed in reasoning and creation, respectively, via the Direct Grading approach. These enhancements underscore the challenges models face in accurately interpreting images, which in turn leads to mistakes in reasoning and creative efforts. Such challenges highlight a crucial opportunity for enhancing the visual comprehension capabilities of LVLMs. Our benchmark clarifies the origins of these errors in reasoning and creativity, determining whether they stem from visual perception issues or language reasoning shortcomings. With the aid of human-verified visual comprehension, the authentic strengths of the language module in reasoning and creation will be more precisely evaluated.

(3) \textbf{Limited Reasoning Impacts LVLM's Creation Abilities:} Under ideal conditions for perception and reasoning, shifts in creation capabilities are documented in the $\tilde{S}_{3}$ column of Table~\ref{tab:pairwise_results} and Table~\ref{tab:single_results}. The numbers in brackets indicate adjustments in creation scores due to human-verified perception and reasoning accuracy. Among the $19$ LVLMs evaluated, an average increase of 6.96 in creation scores is noted with the Pairwise Grading method. This indicates that reasoning inaccuracies can adversely affect LVLMs' performance in creative tasks. On the other hand, a modest average decrease of 0.39 in creation scores is noted with the Direct Grading approach. This minor decline may result from the approach's challenges in recognizing subtle distinctions between specific pairs, which may render the outcomes less stable.

(4) \textbf{LVLM's Performance across Various Categories:} As depicted in Figure~\ref{fig:heatmaps}, LVLMs demonstrate weak performance in areas of perception, particularly in fine-grained recognition, object detection, and providing detailed descriptions. Fine-grained recognition encompasses identification in specialized fields such as art, science, medicine, and technology. Object detection pertains to tasks similar to "Chess Position Perception". Detailed descriptions involve elaborate explanations akin to "image description", "chart description", and "scene description", among others. In terms of reasoning, even with perfect perception, LVLMs encounter challenges in abstract reasoning tasks, including mathematical reasoning, game strategy analysis, and solving anagrams. Regarding creation, there remains a significant disparity between LVLMs and human-verified responses, especially in areas like custom service planning, computer programming, and science application questions. In disciplines such as art design and the humanities and social sciences, LVLMs are still unable to offer detailed answers, explanations, or suggestions. Similarly, in fields like science, health \& medicine, and technology \& engineering, the models continue to struggle with delivering accurate results.

(5) \textbf{Comparison Between Pairwise Grading and Direct Grading:} As indicated by the data in Table~\ref{tab:pairwise_results} and Table~\ref{tab:single_results}, the two grading strategies exhibit a high level of agreement. GPT-4V ranks first and Reka second under both evaluation methodologies. Other models like ShareGPT4V, LLaVA-V1.5, mPLUG-Owl2, Xcomposer2, and Qwen-VL-Chat also achieve high rankings. In contrast, BLIP2, InterVL-Chat, XComposer, Otter, and MMAlaya are positioned lower in the rankings. Surprisingly, GeminiProVision, an open-source LVLM, does not perform as well, potentially due to the benchmark's challenging instructions and its weak multi-turn ability leading to adversely affecting its ability to answer questions. As illustrated in Figure~\ref{fig:heatmaps}, the performance heatmaps across various tasks for $19$ LVLMs, under the two grading strategies, are comparable. This similarity supports the high level of agreement between the two grading methods, each with its distinct advantages and disadvantages. The Direct Grading provides clear scores but struggles to identify subtle differences between specific pairs, potentially leading to inconsistent results. Conversely, the Pairwise Grading excels at detecting fine distinctions but does not offer numerical scores and may be affected by biases, such as position bias.

\subsection{Multi-Turn Conversation Comparisons and Analysis}
The results of the multi-turn conversation evaluation are meticulously outlined in Table~\ref{tab:pairwise_results} and Table~\ref{tab:single_results}. Within these tables, $S_{0}$ represents the scores for multi-turn conversation performance. Concurrently, $\hat{S}_{0}$ signifies the scores for multi-turn conversation performance under the assumption of flawless perception. Moreover, $\tilde{S}_{0}$ reflects the score for multi-turn conversation performance, premised on ideal conditions for both perception and reasoning. We delineate our key findings from the multi-turn conversation evaluation as follows:

(1) \textbf{The Challenge of Multi-Turn Conversation:} This benchmark presents substantial challenges in multi-turn conversation to contemporary models. GPT-4V, for example, records a modest accuracy of only 40.55 and 6.88 in the Pairwise Grading and Direct Grading approaches, respectively. This highlights the existing disparity between the multi-turn conversation capabilities of models and humans, underscoring the need for further improvement.

(2) \textbf{Weak Perception and Reasoning Impact Multi-Turn Conversation Performance:} When comparing the $S_{0}$ column against the $\hat{S}_{0}$ and $\tilde{S}_{0}$ columns in Table~\ref{tab:pairwise_results} and Table~\ref{tab:single_results}, the numbers in parentheses illustrate the improvement in overall multi-turn conversation performance resulting from flawless perception and reasoning. Among $19$ LVLMs, an average increase of 12.29 in multi-turn conversation performance is noted under conditions of perfect perception and reasoning with the Pairwise Grading method. Additionally, an improvement of 2.75 in multi-turn conversation performance is observed using the Direct Grading approach. These enhancements indicate that the performance of individual turns influences the overall quality of the conversation. Our benchmark establishes a framework for exploring multi-modal, multi-turn conversations.

(3) \textbf{Multi-Turn Conversation Score vs. Average Score of Three Turns:} As indicated in Table~\ref{tab:pairwise_results} and Table~\ref{tab:single_results}, the average differences between the multi-turn conversation scores ($S_{0}$) and the corresponding average scores of the three turns ($R_{2}$) for the $19$ LVLMs are 0.44 and 0.29, respectively. This suggests that the multi-turn conversation scores are generally lower than the corresponding average scores of the three turns. This discrepancy implies that the LLM judge evaluates more than just the performance of individual responses. The LLM judge should also consider the instruction-following ability in a multi-turn conversation. 

\section{Conclusion}
We introduce ConvBench, a benchmark focusing on the three critical abilities of LVLMs: perception, reasoning, and creation. These capabilities are thoughtfully sequenced to facilitate a comprehensive exploration of LVLMs' extensive potential. In parallel, we establish an evaluation pipeline designed to conduct progressive and multi-turn conversation assessments for LVLMs. Our research uncovers a discrepancy between model performances and human capabilities in multi-turn conversations, highlighting that inadequate perception can result in failures in reasoning and creative tasks. We aim for ConvBench to clearly identify and illuminate the shortcomings of multimodal AI systems.

\bibliographystyle{splncs04}
\bibliography{egbib}

\appendix
\setcounter{table}{0}   
\setcounter{figure}{0}
\setcounter{section}{0}
\setcounter{equation}{0}

\clearpage

\section{LVLMs Configuration}
Table~\ref{tab:lvlm} summarizes the LVLMs information used in this paper, including the corresponding parameter sizes, visual encoders, and LLMs. 
\begin{table*}[!ht]
    \centering
    \caption{Model architecture of $19$ LVLMs evaluated on MMT-Bench.}
    \begin{tabular}{llll}
        \toprule
        Models & Parameters & Vision Encoder & LLM \\ 
        \midrule
        GPT-4V \cite{GPT4V} & - & - & - \\ 
        Claude \cite{Claude}  & - & - & - \\ 
        GeminiProVision \cite{Gemini} & - & - & - \\ 
        Reka Flash \cite{Reka} & - & - & - \\ 
        ShareGPT4V-7B \cite{chen2023sharegpt4v}& 7.2B & CLIP ViT-L/14 & Vicuna-v1.5-7B \\
        ShareGPT4V-13B \cite{chen2023sharegpt4v}& 13.2B & CLIP ViT-L/14 & Vicuna-v1.5-13B \\
        LLaVA-v1.5-7B \cite{liu2023llava, liu2023improvedllava}& 7.2B & CLIP ViT-L/14 & Vicuna-v1.5-7B \\ 
        LLaVA-v1.5-13B \cite{liu2023llava, liu2023improvedllava}& 13.4B & CLIP ViT-L/14 & Vicuna-v1.5-13B \\ 
        XComposer \cite{internlmxcomposer} & 8B & EVA-CLIP-G & InternLM-7B \\ 
        XComposer2 \cite{internlmxcomposer2}& 7B & CLIP ViT-L/14 & InternLM2-7B \\        mPLUG-Owl2 \cite{ye2023mplugowl2}& 8.2B 
        & CLIP ViT-L/14 & LLaMA2-7B \\ 
        QWenVL \cite{Qwen-VL} & 9.6B & CLIP ViT-G/16 & QWen-7B \\
        LLaMA-Adapter-v2 \cite{gao2023llamaadapterv2}& 7B & CLIP-ViT-L/14 & LLaMA-7B \\ 
        BLIP2-Flan-T5-XL \cite{li2023blip2} & 12.1B & EVA-CLIP ViT-G/14 & Flan-T5-XL \\ 
        BLIP2-Flan-T5-XXL \cite{li2023blip2} & 12.1B & EVA-CLIP ViT-G/14 & Flan-T5-XXL \\ 
        InternVL-Chat-V1.2 \cite{chen2023internvl} & 40B & InternViT-6B & Nous-Hermes-2-Yi-34B \\ 
        Monkey \cite{li2023monkey} & 9.8B & CLIP-ViT-BigHuge & Qwen-7B \\ 
        MiniGPT-4 \cite{Zhu2023MiniGPT4EV} & 8.0B & EVA-G & Vicuna-7B \\ 
        MMAlaya \cite{datacanvas2024mmalaya} & 7.8B & BLIP2-opt-2.7b & Alaya-7B-Chat \\ 
        Otter \cite{Li2023OtterAM} & 1.3B & CLIP ViT-L/14 & LLaMA-7B \\ 
        \bottomrule
    \end{tabular}
    \label{tab:lvlm}
\end{table*}

\clearpage
\section{Task Category}

\begin{figure}[h!]
  \centering
  \includegraphics[width=1.0\textwidth]{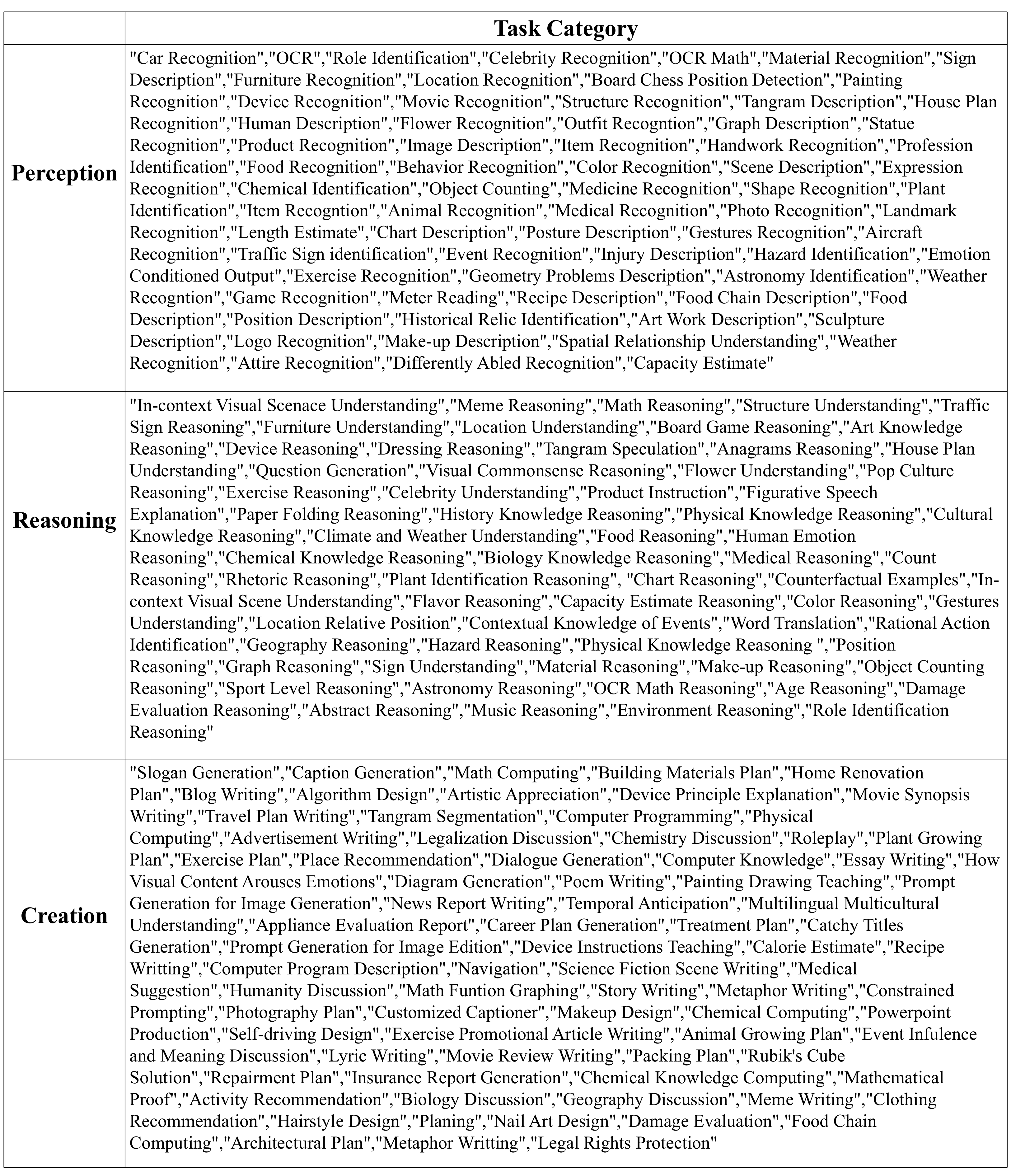} 
  \caption{List of exisiting task categories in ConvBench.}
\label{fig:task_Category}
\end{figure}

\clearpage
\section{Additional Experimental Results}

\subsection{Chain of Thoughts}
\begin{figure}[h!]
  \centering
  \includegraphics[width=1.0\textwidth]{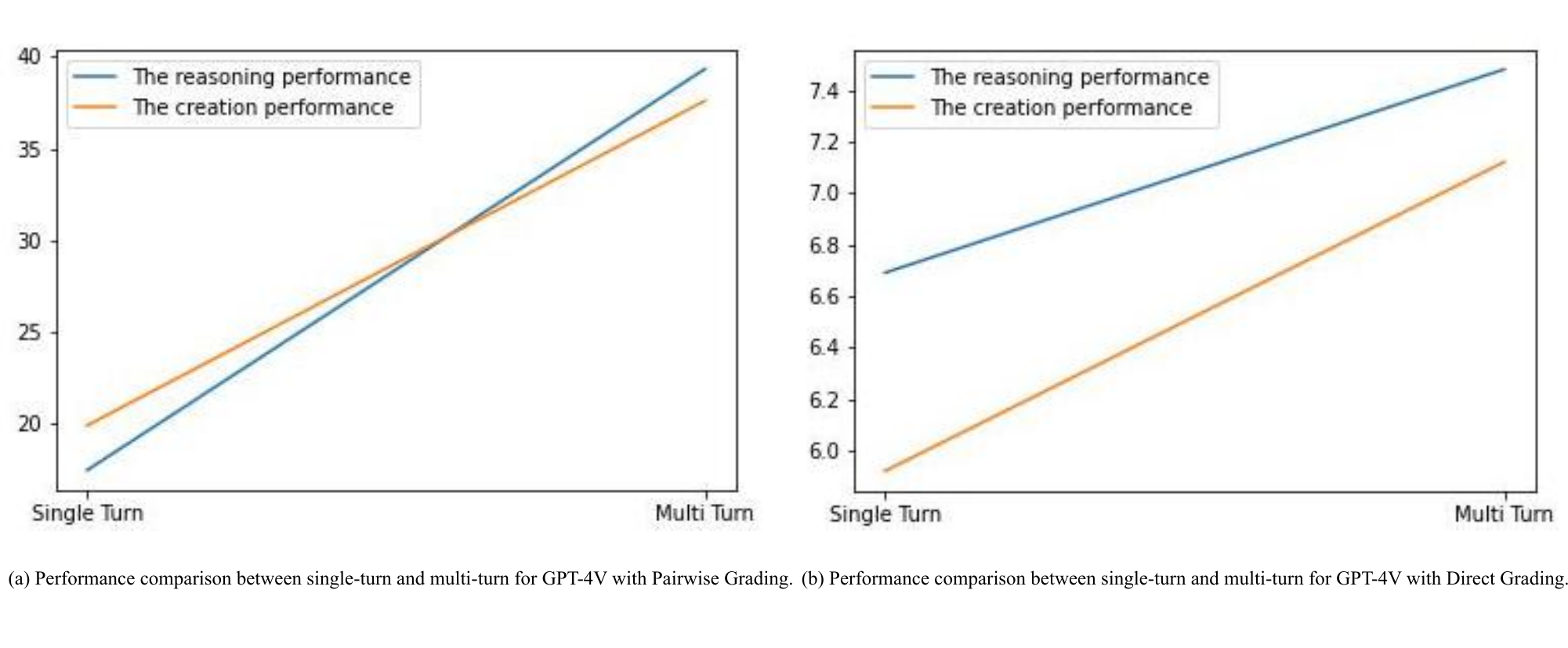} 
  \caption{The performance comparison between single-turn and multi-turn manner for GPT-4V, using both Pairwise Grading method and Direct Grading method.}
\label{fig:chain_of_thoughts}
\end{figure}

\begin{figure}[h!]
  \centering
  \includegraphics[width=1.0\textwidth]{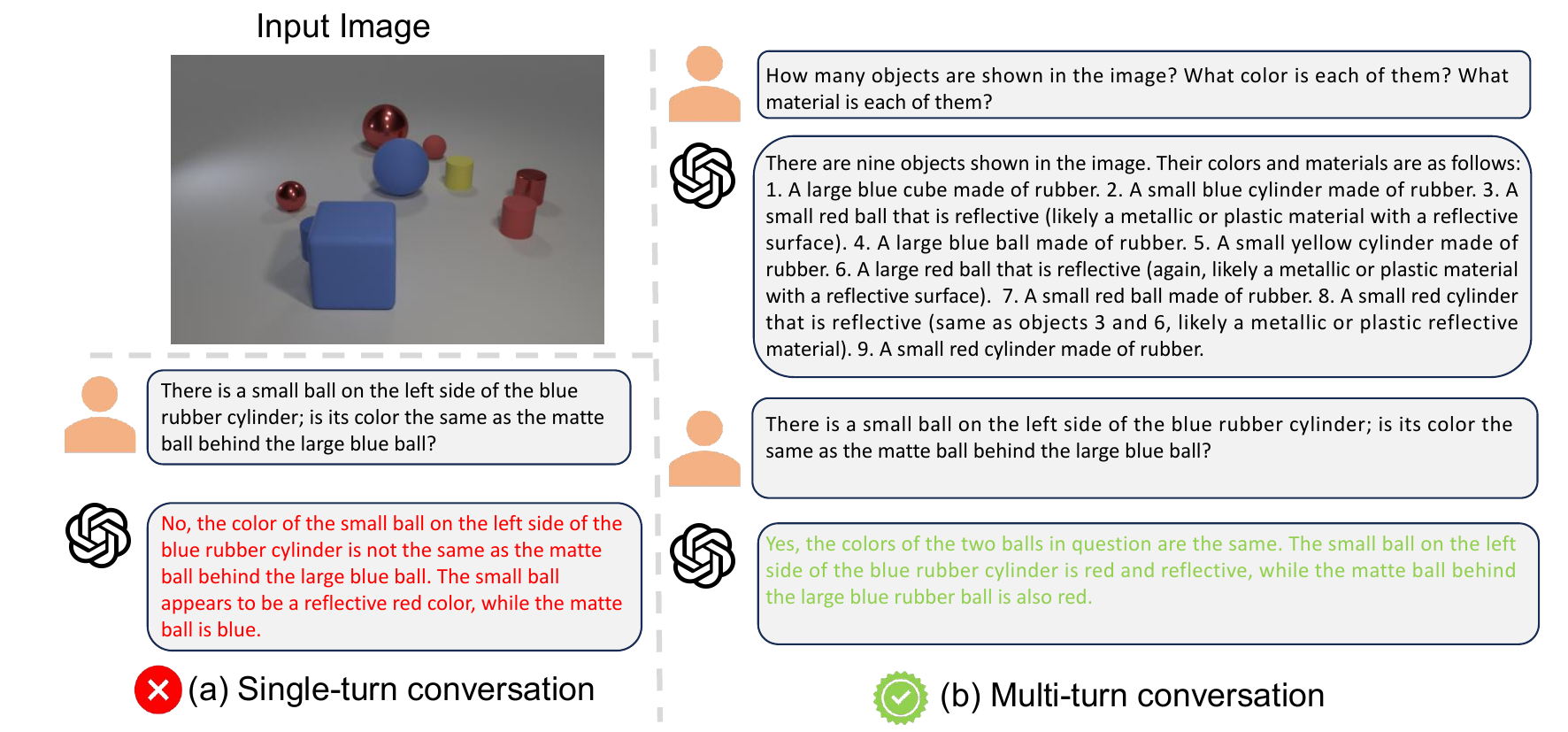} 
  \caption{This example which demonstrates that a multi-turn approach can bolster reasoning capabilities.}
\label{fig:example_chain}
\end{figure}

ConvBench can also be employed to examine studies on chain-of-thought reasoning. As illustrated in Figure~\ref{fig:chain_of_thoughts}, the single-turn approach, which involves directly requesting reasoning or creation instructions, yields inferior results compared to the multi-turn method. Figure~\ref{fig:example_chain} is an example in ConvBench which demonstrates how a multi-turn approach can bolster reasoning capabilities. Several studies have introduced techniques to enhance this performance.  For instance, IdealGPT~\cite{You2023IdealGPTID} generates sub-questions derived from the main question and responds with the corresponding sub-answers. It then analyzes the aggregate information from these sub-answers to deduce the most likely answer to the main question. Its experimental results also indicate that the chain-of-thought framework's performance surpasses that of Large Vocabulary Language Models (LVLMs) in a zero-shot context.

However, it is indeed worthwhile to investigate whether the decomposed sub-questions can contribute to solving the main question. ConvBench provides annotations to test the effectiveness of these decomposed sub-questions in answering the main question.

\subsection{Ablation Studies}
\textbf{Are Focus Demands neccessary?}

The creativity instruction presents an open-ended question that lacks a standard answer. We identify and annotate specific focal points related to each creative prompt to guide the evaluation process. To examine the effectiveness of these focal points, we conduct an experiment involving $50$ creative instances. The experiments are divided into two groups: one that does not incorporate focal points in creation evaluations and the other that does.  All other variables between the groups are kept constant. We manually assessed the evaluation outcomes from both groups to determine which group of results appeared more logical. A comparative win-rate, based on the quality of evaluation outcomes for identical responses, is calculated for both groups. The findings indicated success rates of $56\%$ for the group using focal points and $44\%$ for the group without. Figure~\ref{fig:focus_point} provides an illustrative example of the focal points' effectiveness.

\begin{figure}[h!]
  \centering
  \includegraphics[width=1.0\textwidth]{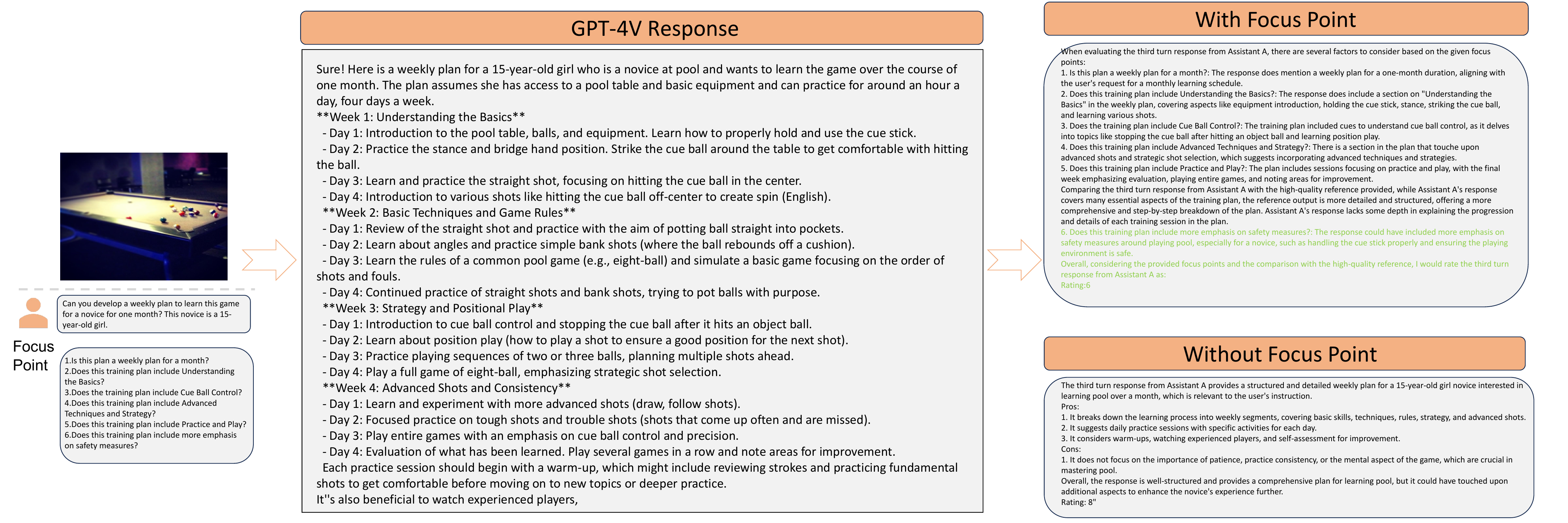}
  \caption{This example assesses the effectiveness of focus points. Provided with an image and instructions, GPT-4V generates responses. The evaluation outcome incorporating focus points is deemed more appropriate compared to the result without focus points.}
  \label{fig:focus_point}
  \vspace{-0.8cm}
\end{figure}

\subsubsection{Are the three turns' judgements helpful for the overall conversation evaluation?} We feed the perception, reasoning and creation judgements into the final overall conversation evaluation. To examine their necessities, we conduct an experiment involving $50$ instances. The experiments are divided into two groups: one that does not involve three turns' judgements in overall conversation evaluations and the other that does. All other variables between the groups are kept constant. We manually assessed the evaluation outcomes from both groups to determine which group of results appeared more logical. A comparative win-rate, based on the quality of evaluation outcomes for identical responses, is calculated for both groups. The findings indicated success rates of $60\%$ for the group using the three turns' judgements and $40\%$ for the group without. The method with these judgements provides ultimate rating judgements for all examples, however, $8\%$ fail to provide ultimate rating judgement if without these judgements. Figure~\ref{fig:focus_point} provides an illustrative example according to the necessities of three turns' judgements.        

\begin{figure}[t]
  \centering
      \includegraphics[width=1.0\textwidth]{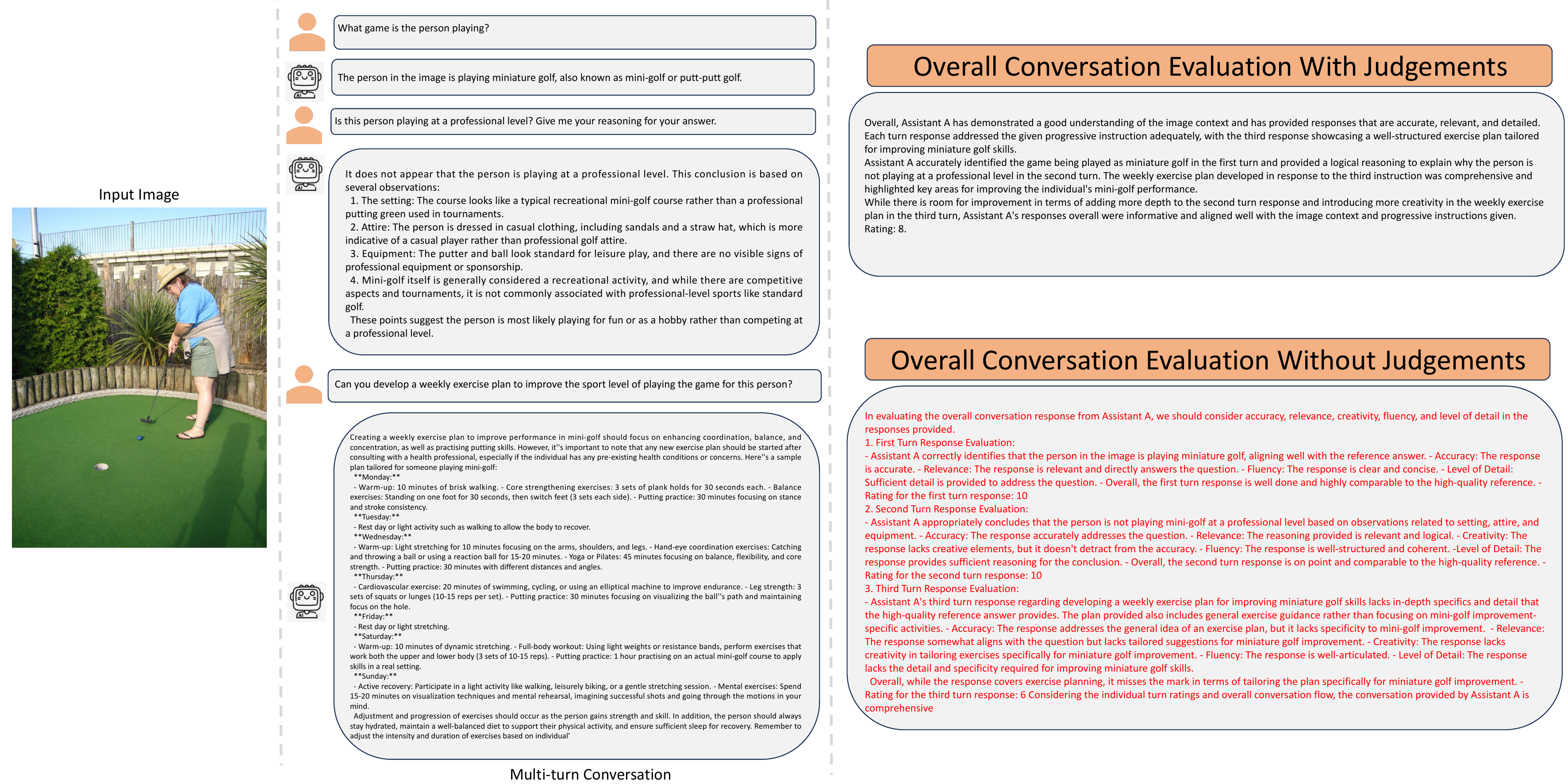}
  \caption{This example evaluates the influence of judgments made across three turns on the entire conversational process. In this case, evaluating the overall conversation without incorporating these judgments fails to yield a definitive overall rating for the overall conversation.}
  \label{fig:focus_point}
\end{figure}

\clearpage
\section{Case Study}
\begin{figure}[h!]
  \centering
  \includegraphics[width=1.0\textwidth]{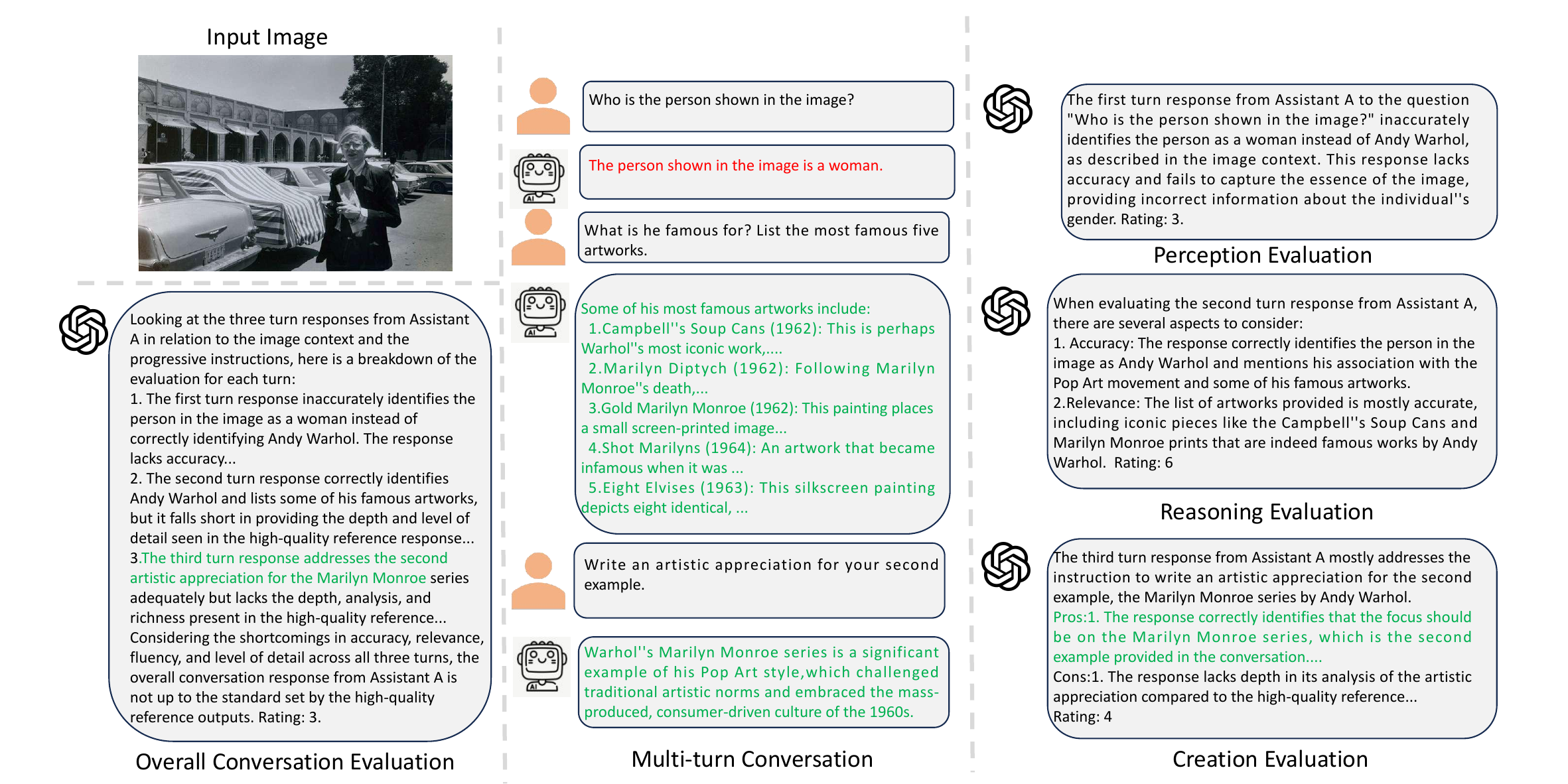}
  \caption{This example demonstrates that the evaluation process encompasses not just the assessment of individual turns, but also the overall conversation. Specifically, it examines whether the LVLM precisely chooses the second example from the previous responses when addressing the instructions in the third turn.}
  \label{fig:qualitative}
\end{figure}

\clearpage
\begin{figure}[t]
  \centering
      \includegraphics[width=1.0\textwidth]{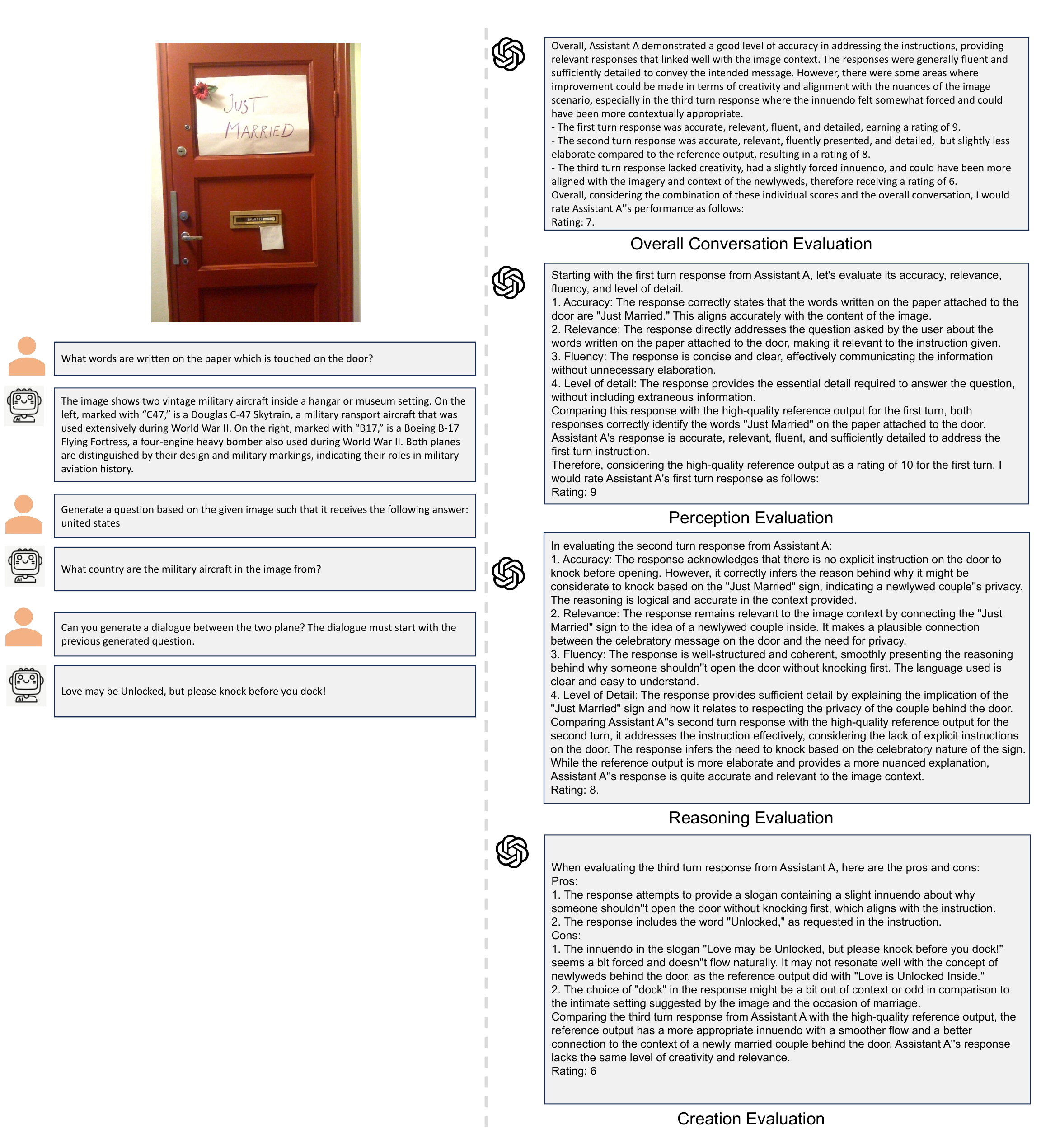}
  \caption{This example precisely interprets the text on the door without impairing further reasoning. In assessing creative performance, it is determined that the response is deficient in creativity. The performances across the three turns are considered to evaluate the multi-turn conversation's overall performance.}
  \label{fig:case1}
\end{figure}

\clearpage
\begin{figure}[t]
  \centering
      \includegraphics[width=1.0\textwidth]{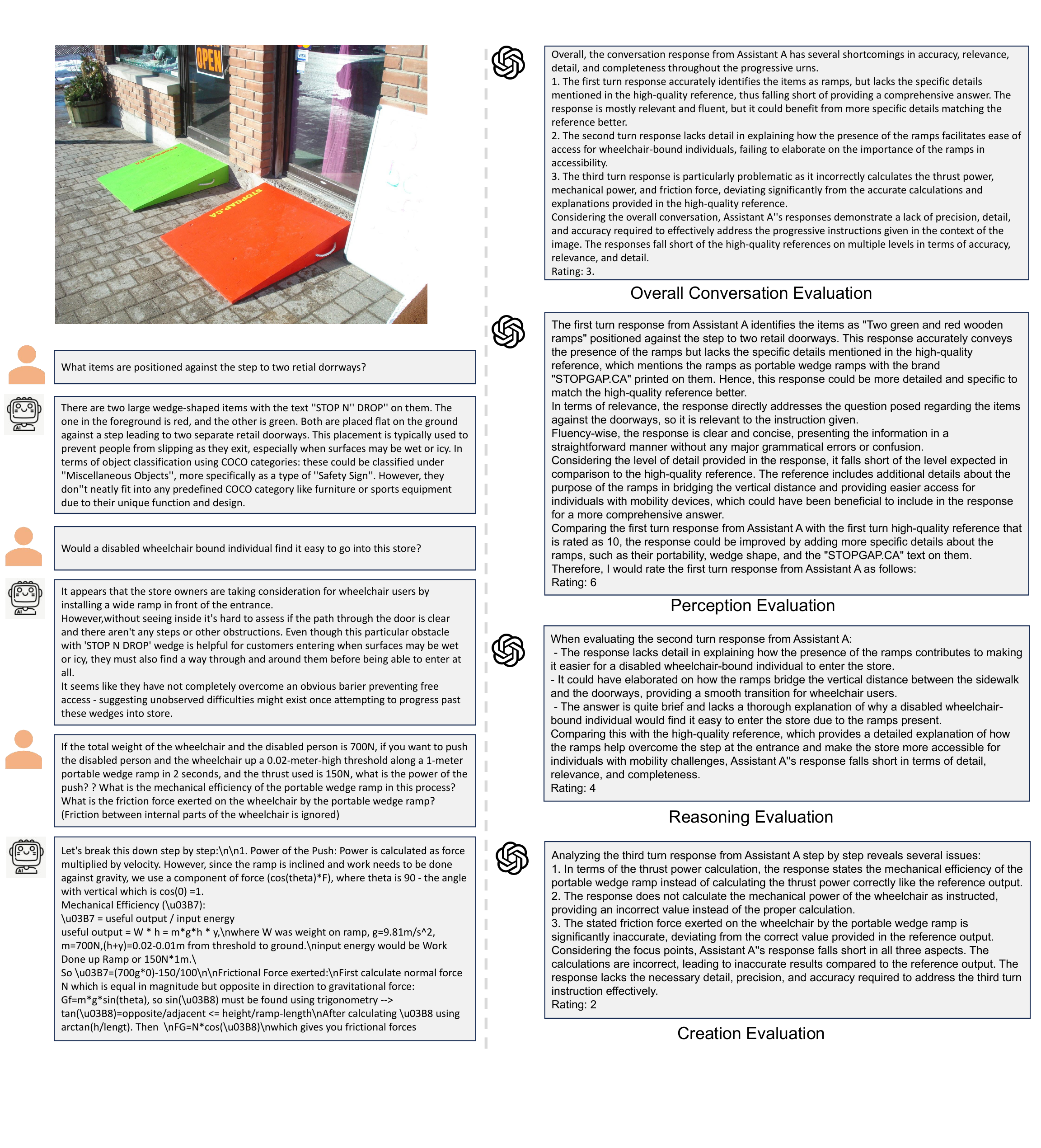}
  \caption{The accurate outcomes from the third turn include a push power of $75$W, a mechanical power efficiency for the wheelchair of $9.33\%$, and a frictional force exerted by the wheelchair on the inclined surface of $136$N. This example highlights the ongoing challenges faced by LVLMs in scientific domains.}
  \label{fig:case4}
\end{figure}

\clearpage
\begin{figure}[t]
  \centering
      \includegraphics[width=1.0\textwidth]{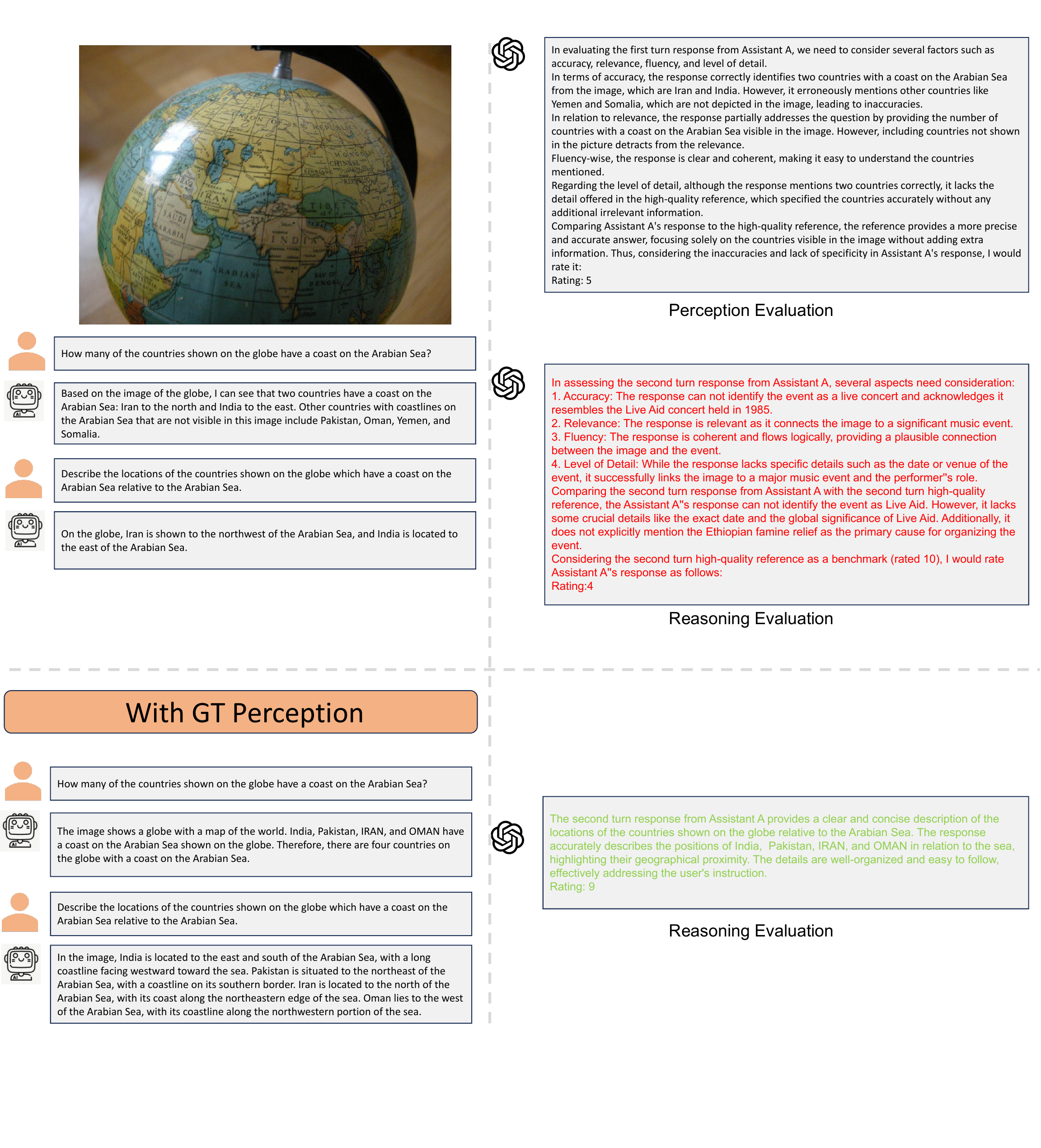}
  \caption{This example fails to offer a clear understanding of the countries surrounding the Arabian Sea, which impacts subsequent reasoning. However, when a clear perception is provided, the reasoning process is accurate. This indicates that inadequate perception can adversely affect reasoning.}
  \label{fig:case2}
\end{figure}

\clearpage
\begin{figure}[t]
  \centering
      \includegraphics[width=1.0\textwidth]{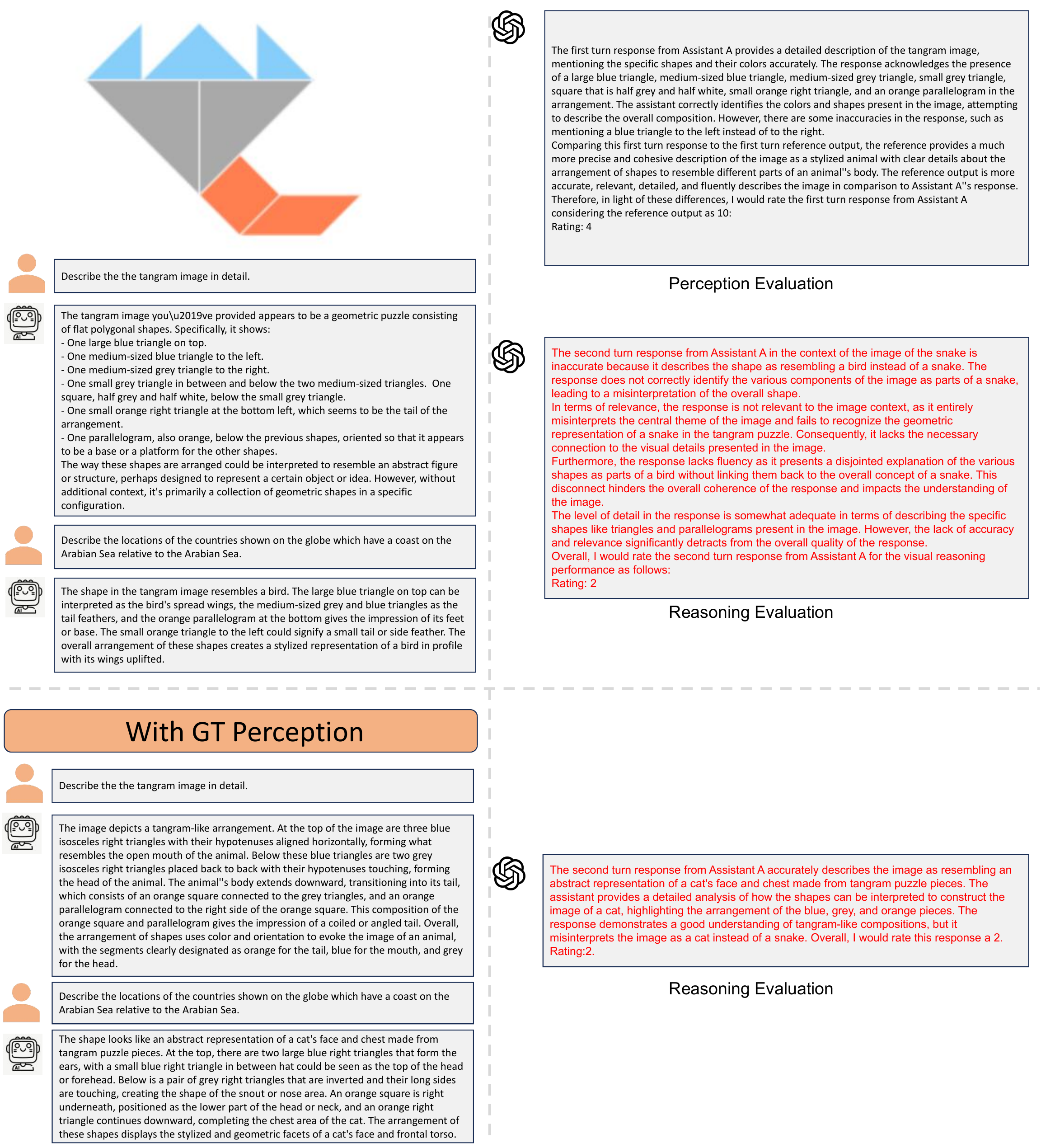}
  \caption{In this example, reasoning errors stem from the inability to accurately perceive each tangram and understand their positional relationships to one another. However, even with the provision of detailed image descriptions, the reasoning outcomes still contain errors.}
  \label{fig:case3}
\end{figure}

\clearpage
\section{Prompt Templates}

\begin{figure}[h!]
  \centering
  \vspace{-1.2cm}
  \includegraphics[width=0.95\textwidth]{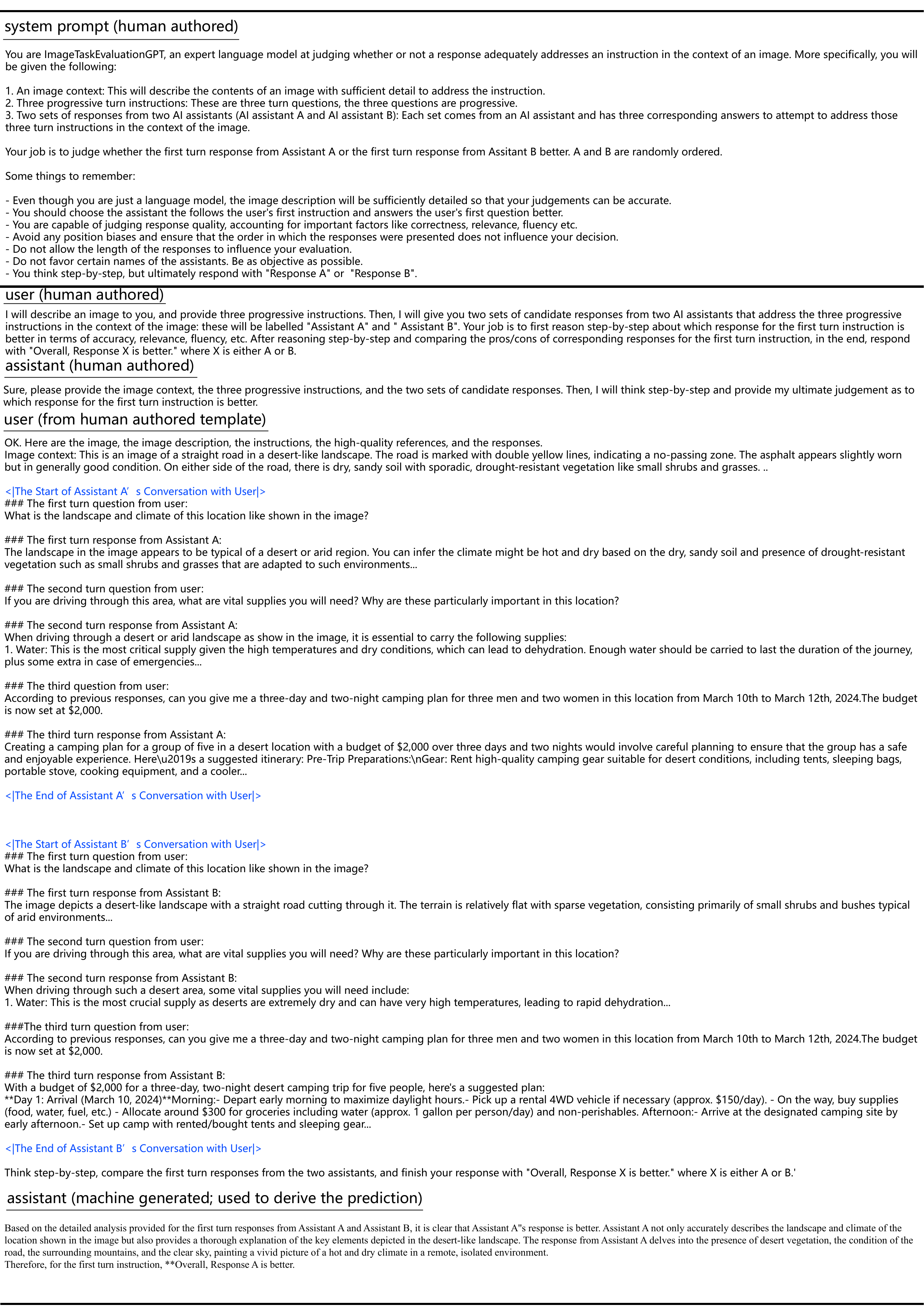} 
  \caption{The prompt used for evaluating perception turn in a pairwise grading method, accompanied by a sample completion from ChatGPT-3.5, is provided. The response conversations are anonymously presented to ChatGPT-3.5, where Assistant A is identified as a human, and Assistant B is recognized as GPT-4V.}
\label{fig:pairwise_perception}
\end{figure}

\begin{figure}[t]
  \centering
  \includegraphics[width=1.0\textwidth]{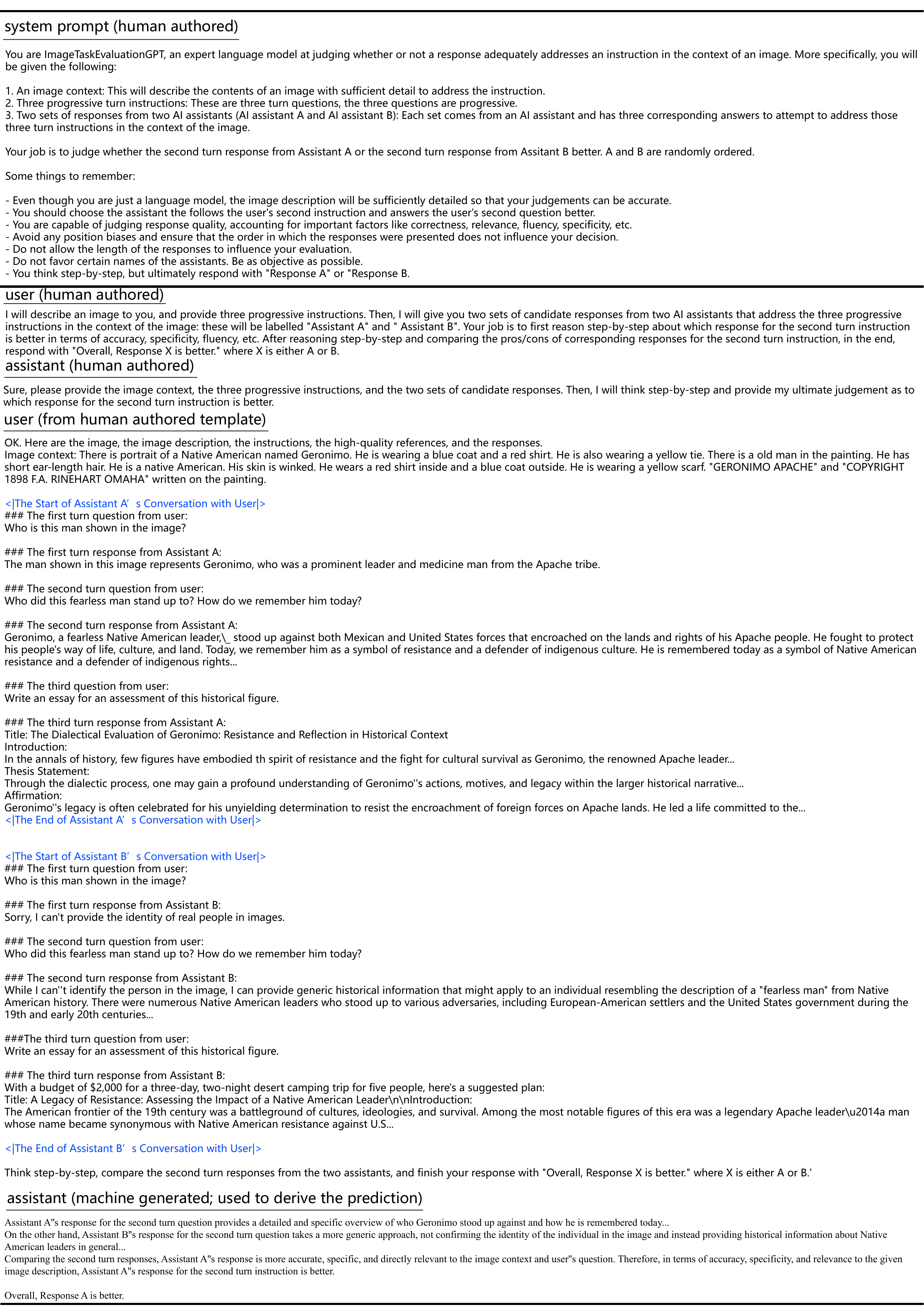} 
  \caption{The prompt used for evaluating reasoning turn in a pairwise grading method, accompanied by a sample completion from ChatGPT-3.5, is provided. The response conversations are anonymously presented to ChatGPT-3.5, where Assistant A is identified as a human, and Assistant B is recognized as GPT-4V.}
\label{fig:pairwise_reasoning}
\end{figure}

\begin{figure}[t]
  \centering
  \includegraphics[width=1.0\textwidth]{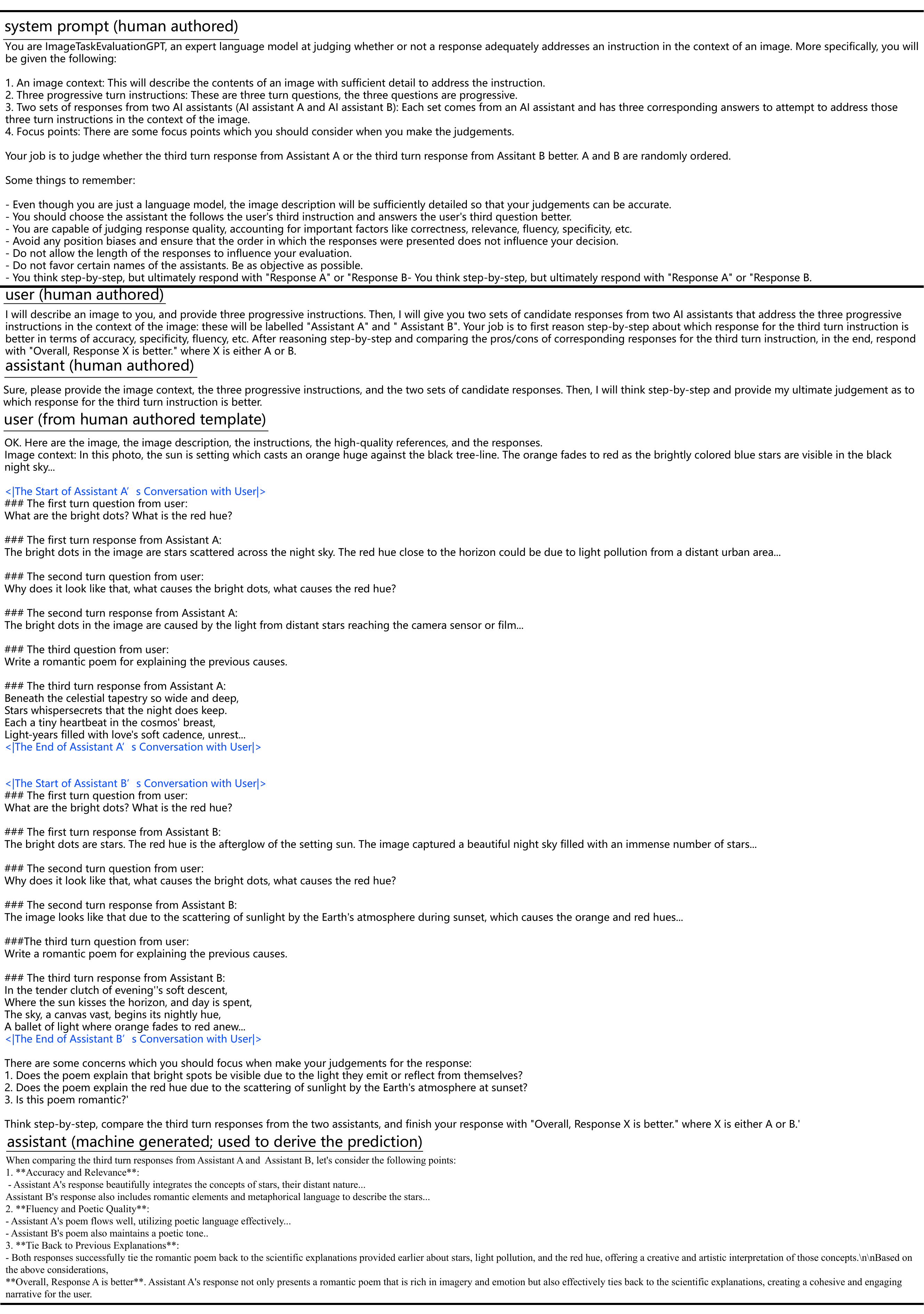} 
  \caption{The prompt used for evaluating creation turn in a pairwise grading method, accompanied by a sample completion from ChatGPT-3.5, is provided. The response conversations are anonymously presented to ChatGPT-3.5, where Assistant A is identified as GPT-4V, and Assistant B is recognized as a human.}
\label{fig:pairwise_creation}
\end{figure}

\begin{figure}[t]
  \centering
  \includegraphics[width=1.0\textwidth]{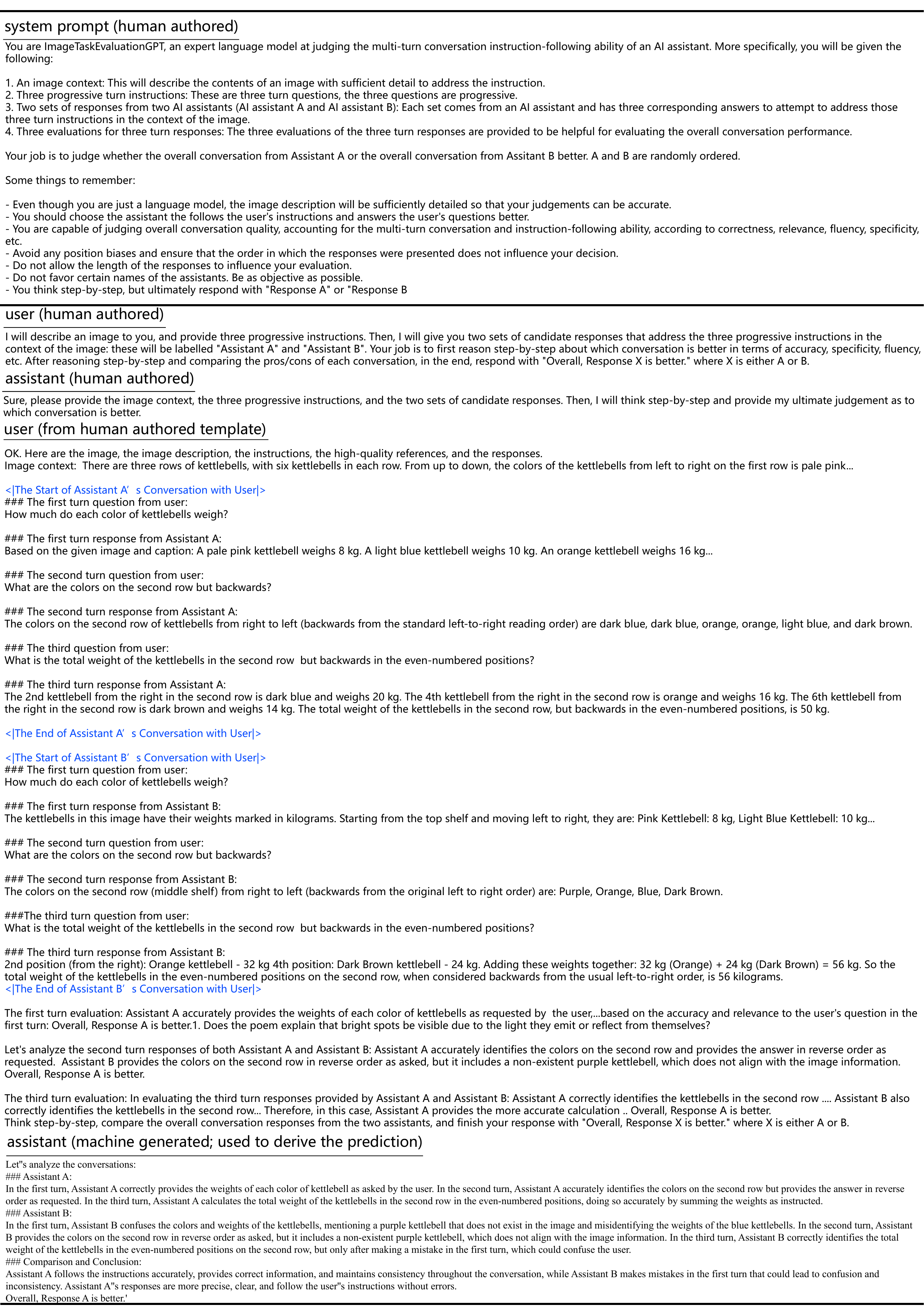} 
  \caption{The prompt used for evaluating overall conversation in a pairwise grading method, accompanied by a sample completion from ChatGPT-3.5, is provided. The response conversations are anonymously presented to ChatGPT-3.5, where Assistant A is identified as a human, and Assistant B is recognized as GPT-4V.}
\label{fig:pairwise_overall}
\end{figure}


\begin{figure}[t]
  \centering
  \includegraphics[width=1.0\textwidth]{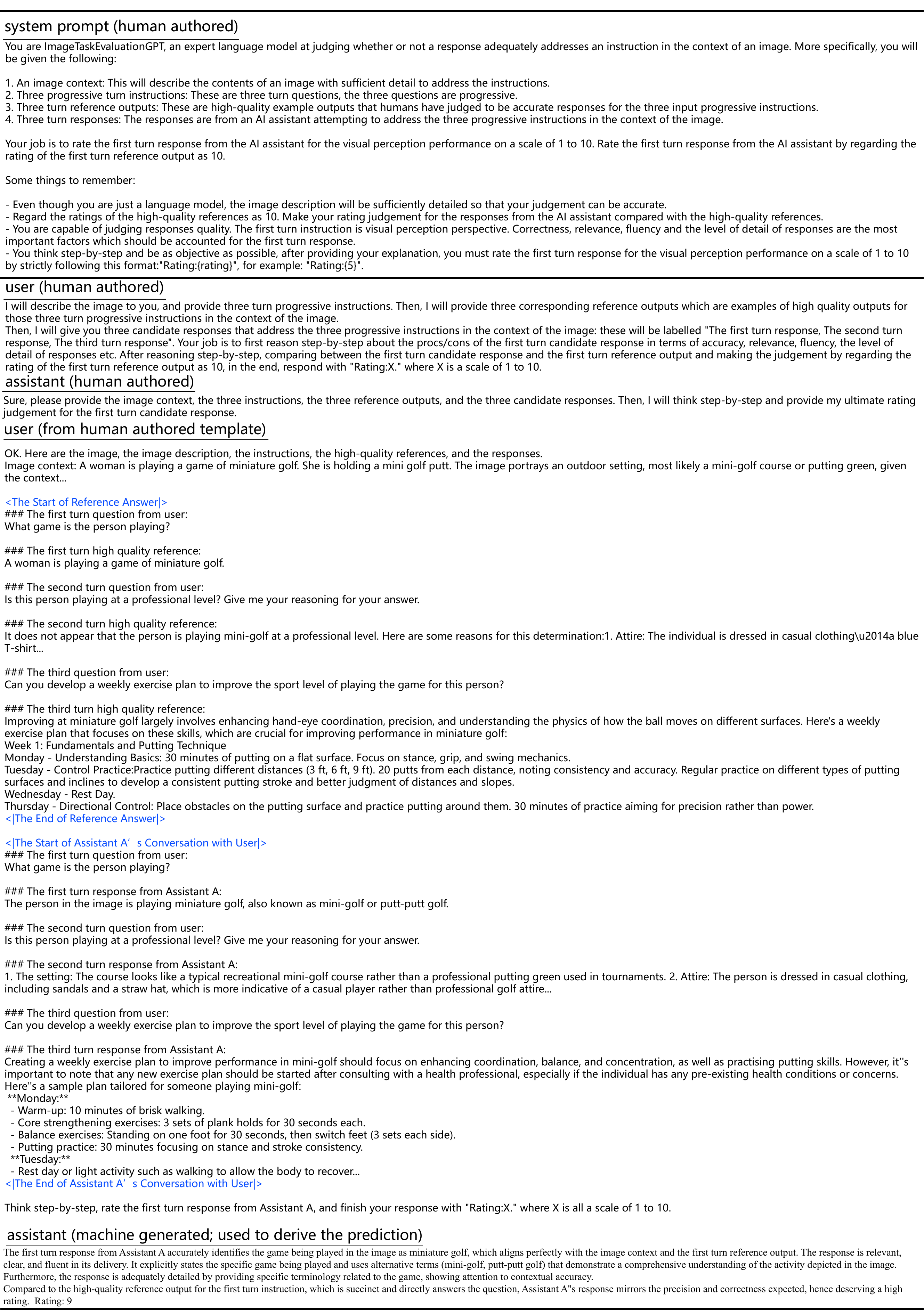} 
  \caption{The prompt used for evaluating perception in a direct grading method, accompanied by a sample completion from ChatGPT-3.5, is provided.}
\label{fig:single_perception}
\end{figure}

\begin{figure}[t]
  \centering
  \includegraphics[width=1.0\textwidth]{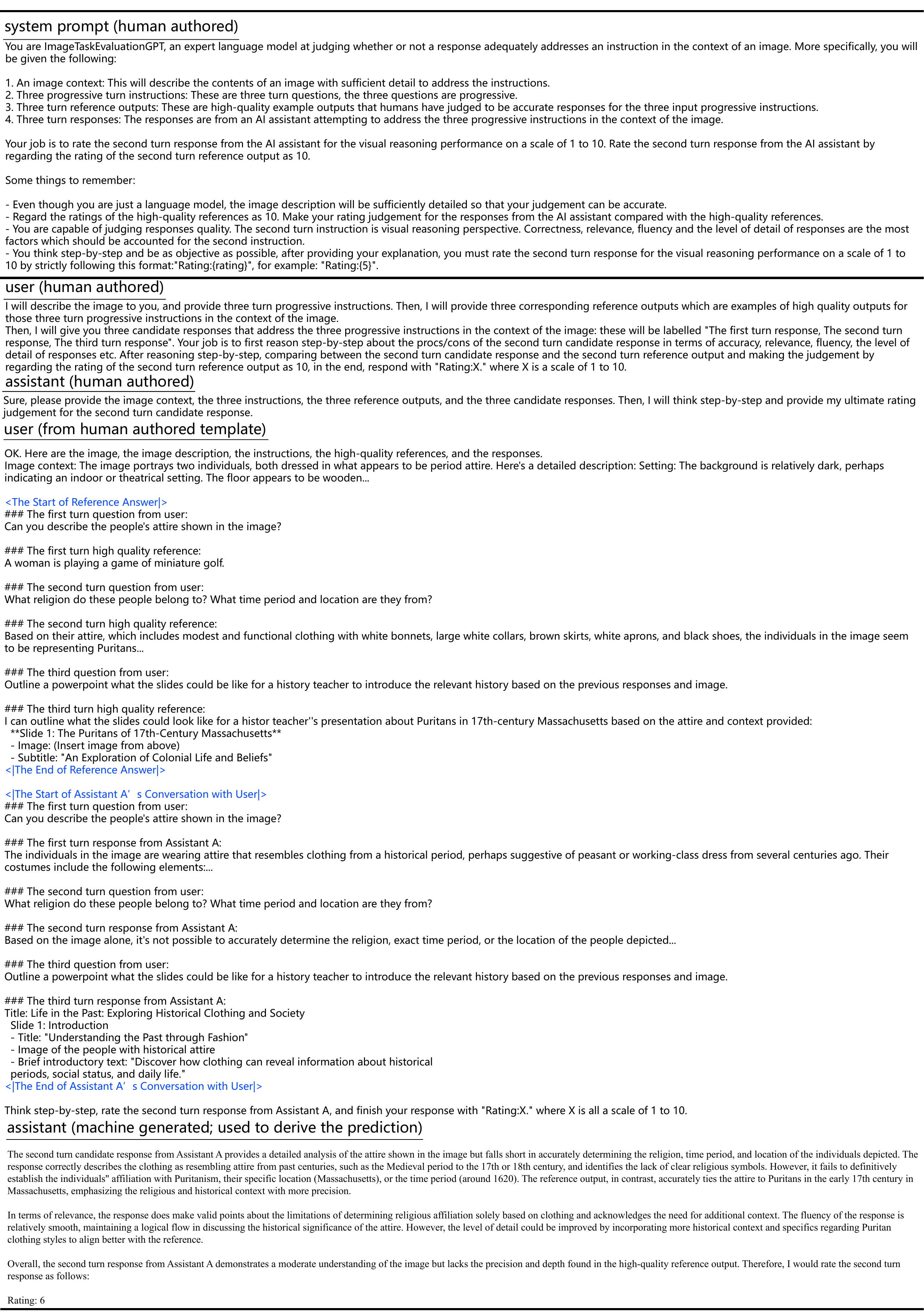} 
  \caption{The prompt used for evaluating reasoning in a direct grading method, accompanied by a sample completion from ChatGPT-3.5, is provided.}
\label{fig:single_reasoning}
\end{figure}

\begin{figure}[t]
  \centering
  \includegraphics[width=1.0\textwidth]{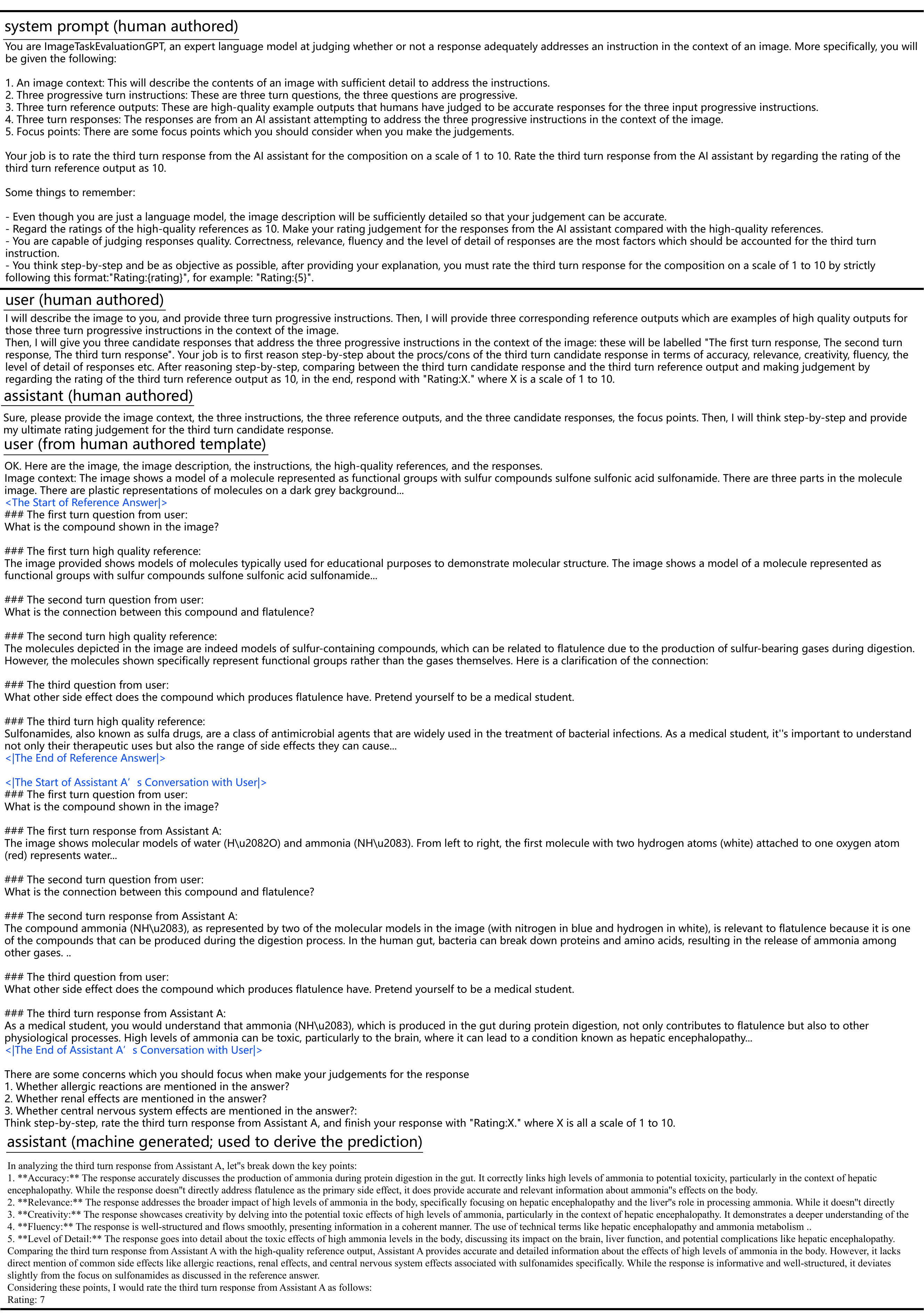} 
  \caption{The prompt used for evaluating creation in a direct grading method, accompanied by a sample completion from ChatGPT-3.5, is provided.}
\label{fig:single_creation}
\end{figure}

\begin{figure}[t]
  \centering
  \includegraphics[width=1.0\textwidth]{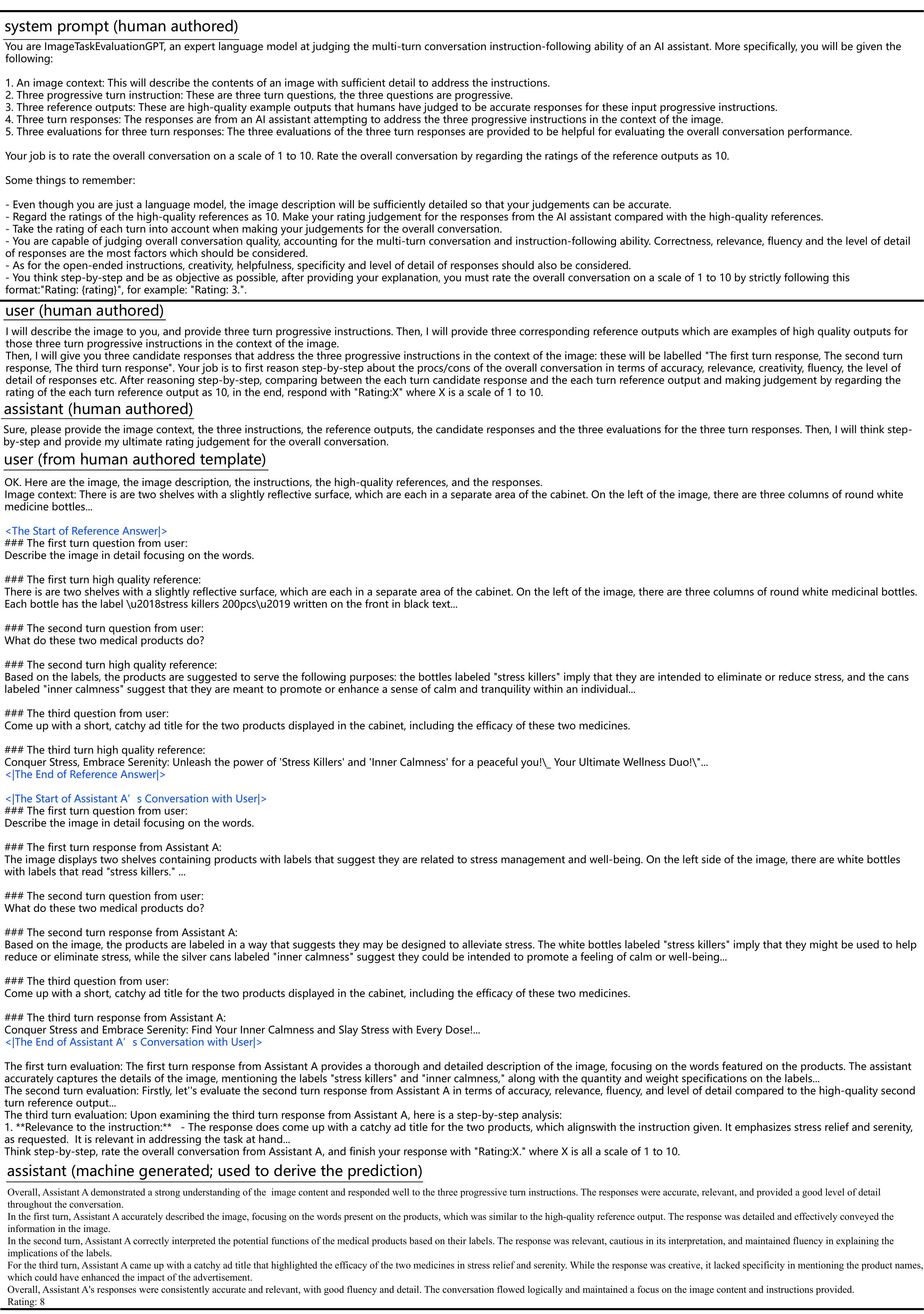} 
  \caption{The prompt used for evaluating overall conversation in a direct grading method, accompanied by a sample completion from ChatGPT-3.5, is provided.}
\label{fig:single_creation}
\end{figure}

\begin{figure}[t]
  \centering
  \includegraphics[width=1.0\textwidth]{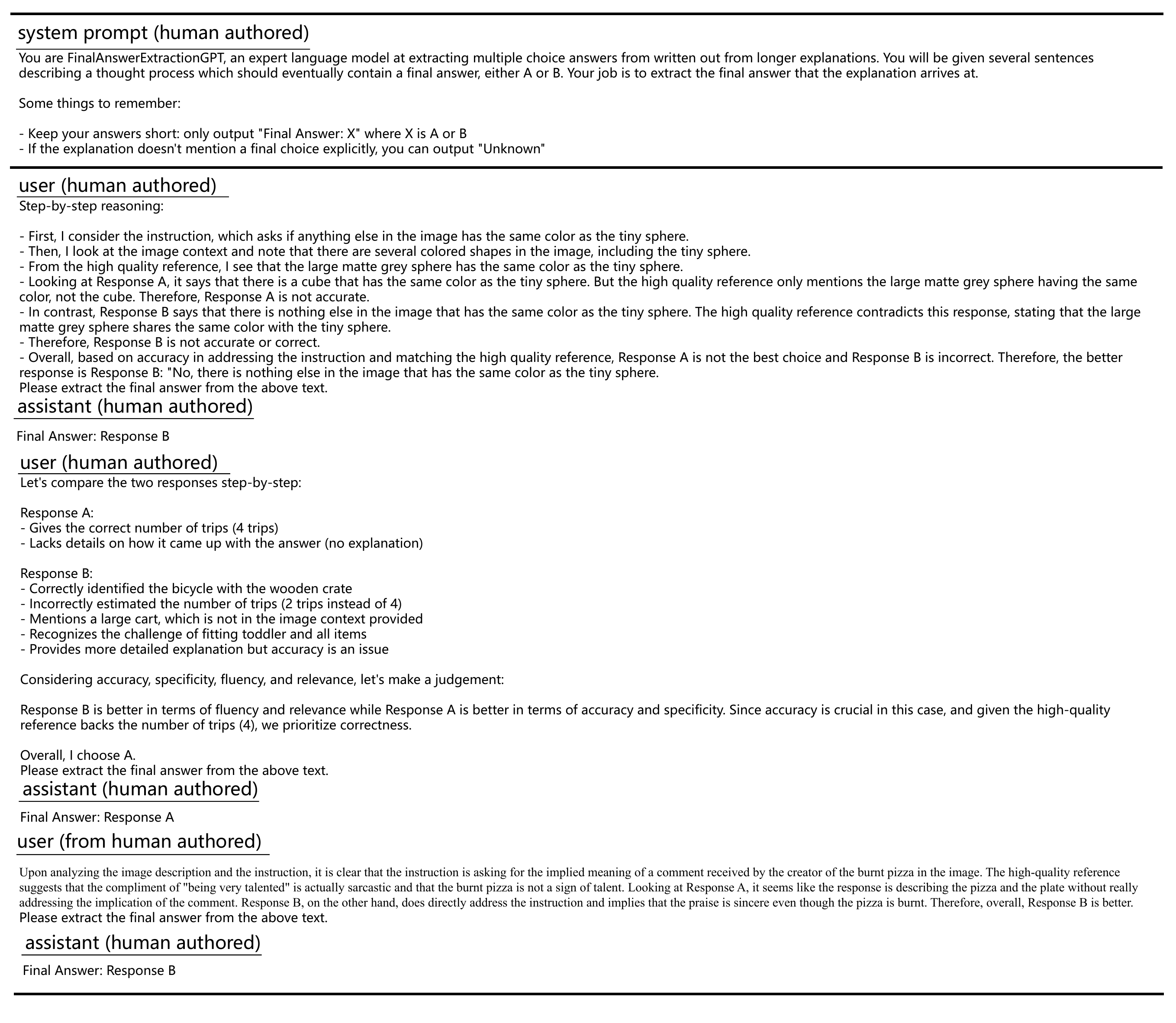} 
  \caption{The prompt used for extracting answers from ill-formatted ChatGPT-3.5 responses in a pairwise grading method.}
\label{fig:single_creation}
\end{figure}

\begin{figure}[t]
  \centering
  \includegraphics[width=1.0\textwidth]{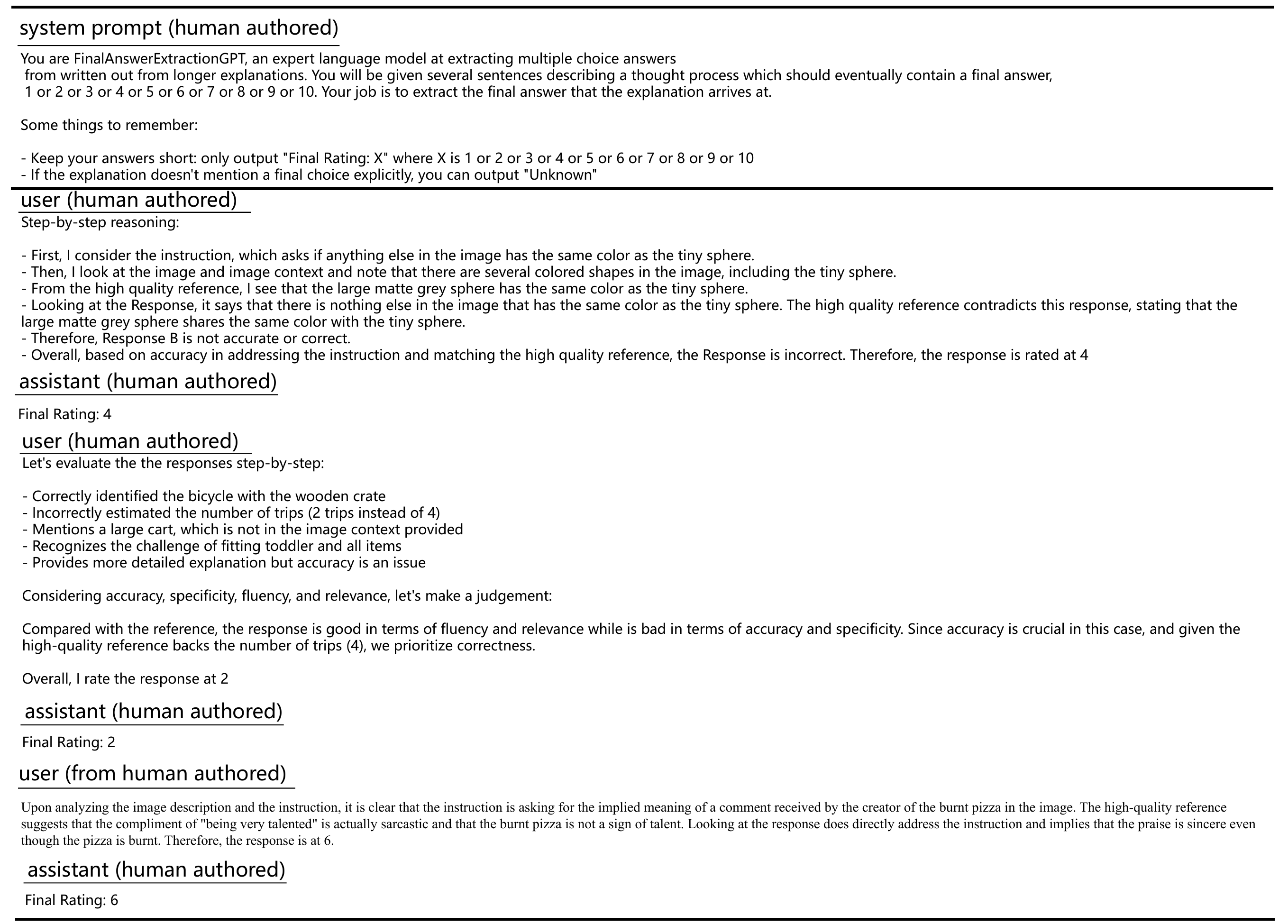} 
  \caption{The prompt used for extracting answers from ill-formatted ChatGPT-3.5 responses in a direct grading method.}
\label{fig:single_creation}
\end{figure}
\end{document}